# DLMP of Competitive Markets in Active Distribution Networks: Models, Solutions, Applications, and Visions

Xiaofei Wang, *Graduate Student Member, IEEE*, Fangxing Li, *Fellow, IEEE*, Linquan Bai, *Senior Member, IEEE*, Xin Fang, *Senior Member, IEEE*

*Abstract*— Traditionally, the electric distribution system operates with uniform energy prices across all system nodes. However, as the adoption of distributed energy resources (DERs) propels a shift from passive to active distribution network (ADN) operation, a distribution-level electricity market has been proposed to manage new complexities efficiently. In addition, distribution locational marginal price (DLMP) has been established in the literature as the primary pricing mechanism. The DLMP inherits the LMP concept in the transmission-level wholesale market, but incorporates characteristics of the distribution system, such as high R/X ratios and power losses, system imbalance, and voltage regulation needs. The DLMP provides a solution that can be essential for competitive market operation in future distribution systems. This paper first provides an overview of the current distribution-level market architectures and their early implementations. Next, the general clearing model, model relaxations, and DLMP formulation are comprehensively reviewed. The state-of-the-art solution methods for distribution market clearing are summarized and categorized into centralized, distributed, and decentralized methods. Then, DLMP applications for the operation and planning of DERs and distribution system operators (DSOs) are discussed in detail. Finally, visions of future research directions and possible barriers and challenges are presented.

*Index Terms* — ACOPF; active distribution network (ADN); centralized, distributed and decentralized solution methods; convexification; DCOPF; demand response; distributed energy resource (DER); distribution-level electricity market; distribution locational marginal price (DLMP); distribution system operator (DSO); linearization; microgrid; P2P trading.

## NOMENCLATURE

*Abbreviations*

| | |
|---|---|
| ACOPF | Alternating Current Optimal Power Flow |
| ADMM | Alternating Direction Method of Multipliers |
| ADN | Active Distribution Network |
| APP | Auxiliary Problem Principle |
| ATC | Analytical Target Cascading |
| BS | Bill Sharing |
| CAISO | California Independent System Operator |
| CHP | Combined Heat and Power |
| CPP | Critical Peak Pricing |
| DCOPF | Direct Current Optimal Power Flow |
| DER | Distributed Energy Resource |
| DG | Distributed Generator |
| DLMP | Distribution Locational Marginal Price |
| DMO | Distribution Market Operator |
| DNO | Distribution Network Operator |
| DR | Demand Response |
| DSO | Distribution System Operator |
| ES | Energy Storage |
| EV | Electric Vehicle |
| FC | Fuel Cell |
| GA | Genetic Algorithm |
| HVAC | Heating, Ventilation, and Air Conditioning |
| ICT | Information and Communications Technology |
| ISO | Independent System Operator |
| LA | Load Aggregator |
| LC | Large Consumer |
| LEM | Local Energy Market |
| LF-D | Loss Factor for the Distribution |
| LFM | Local Flexibility Market |
| LMP | Locational Marginal Price |
| LP | Linear Programming |
| LPF-D | Linearized Power Flow for Distribution |
| MG | Microgrid |
| MILP | Mixed-Integer Linear Programming |
| MMR | Mid-Market Rate |
| MT | Microturbine |
| NB | Nash Bargaining |
| NLP | Nonlinear Programming |
| OPF | Optimal Power Flow |
| P2P | Peer-to-Peer |
| PMP | Proximal Message Passing |
| PSO | Particle Swarm Optimization |
| PTR | Peak Time Rebate |
| PV | Photovoltaic |
| QP | Quadratic Programming |
| RTP | Real-Time Pricing |
| SDP | Semidefinite Programming |
| SOCP | Second-Order Cone Programming |
| TCL | Thermostatically Controlled Load |
| TLBO | Teaching-Learning Based Optimization |
| TOU | Time of Use |

Corresponding author: F. Li.
This work was supported in part by CURENT, an Engineering Research Center funded by the National Science Foundation (NSF) and the Department of Energy under NSF award EEC-1041877, and in part by NSF award ECCS-1809458.

X. Wang and F. Li are with the Dept. of Electrical Engineering and Computer Science, The University of Tennessee, Knoxville, USA.
L. Bai is with University of North Carolina, Charlotte, NC, USA.
X. Fang is with National Renewable Energy Laboratory (NREL), Golden, CO, USA.

| TSO | Transmission System Operator |
| WT | Wind Turbine |

*Sets*

| $\Omega_L$ | Set of distribution lines |
| $\Omega_N$ | Set of nodes |
| $\Omega_G$ | Set of generation resources |

*Variables*

| $P_i^G / Q_i^G$ | Active/reactive power of generator $i$ |
| $P_i^D / Q_i^D$ | Active/reactive load demand of node $i$ |
| $P^{loss} / Q^{loss}$ | Active/reactive power loss |
| $S_l$ | Power flow of line $l$ |
| $V_j$ | Voltage magnitude of bus $j$ |
| $\lambda^p / \lambda^q$ | Lagrange multipliers associated with active/reactive equality power constraints |
| $\omega_l^{s,min} / \omega_l^{s,max}$ | Lagrange multipliers associated with power flow limits |
| $\omega_j^{v,min} / \omega_j^{v,max}$ | Lagrange multipliers associated with voltage limits |
| $\omega_i^{p,min} / \omega_i^{p,max}$ | Lagrange multipliers associated with active power limits |
| $\omega_i^{q,min} / \omega_i^{q,max}$ | Lagrange multipliers associated with reactive power limits |
| $\kappa_i^- / \kappa_i^+$ | Lagrange multipliers associated with inequality reactive power constraints |

*Constants*

| $S_l^{min} / S_l^{max}$ | Minimum/maximum power flow limits |
| $V^{min} / V^{max}$ | Minimum/maximum voltage limits |
| $P_i^{G,min} / P_i^{G,max}$ | Minimum/maximum active power limits |
| $Q_i^{G,min} / Q_i^{G,max}$ | Minimum/maximum reactive power limits |

# I. INTRODUCTION

TRADITIONALLY, power systems are unidirectionally structured, that is, electricity is generated by generators and then transported via transmission and distribution lines to customers, who are at the end of the electricity delivery chain. However, in recent decades, the electricity industry has witnessed the emergence of various distributed energy resources (DERs) in the distribution system. On the supply side, the deployment of distributed generators (DGs), such as photovoltaics (PVs), microturbines (MTs), wind turbines (WTs), energy storage systems (ESSs), etc., in modern distribution systems continues to grow [1][2]. On the demand side, industrial and commercial customers are incentivized and encouraged to participate in demand response (DR) programs. The desire to further expand DR potential, combined with advances in information and communication technology (ICT), has led to residential loads, such as smart home appliances, being widely studied in recent years [3]-[6].

The proliferation of a variety of DERs has transformed the once unidirectional power system into a bidirectional system, making the distribution system more flexible and active, but also more complex. This transition is illustrated in Fig. 1.

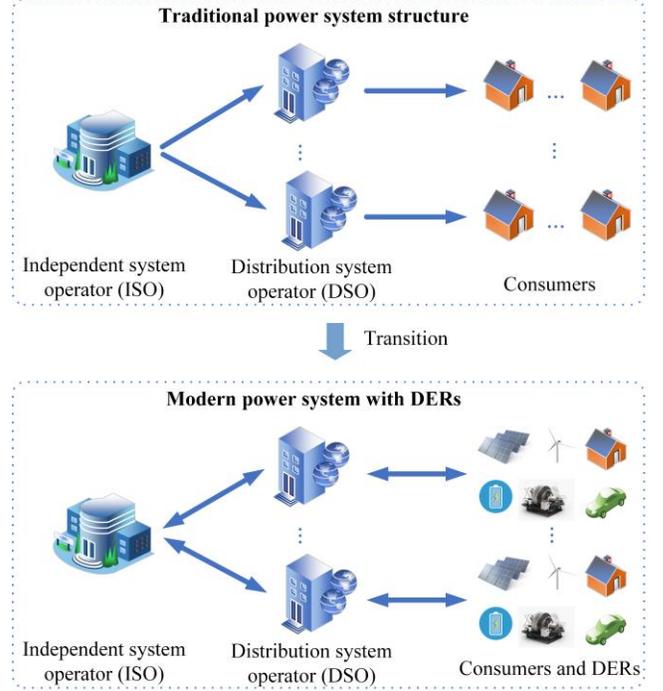

Fig. 1. Diagram of power system transition

## A. Distribution-level Electricity Market

Locally optimizing both the generation and the consumption of electricity can provide many advantages for a power system, such as reducing operating stress in peak load hours, reducing the risk of system failure, and deferring future grid investment. To take advantage of these new opportunities and keep pace with the deregulation process, the distribution system needs to change the traditional business model, market design, and system operation [7]. In the research community, it is believed that a market-based distribution system is a promising means of achieving the optimal allocation of all DERs and improving the system operating efficiency [8]. Various electricity market concepts at the distribution level have been proposed, such as retail markets [9], local flexibility markets (LFM) [10]-[12], local energy markets (LEM) [13]-[15], and transactive energy schemes [16]. In the power industry, there are several ongoing local market pilot projects [17]-[25].

## B. Distribution Locational Marginal Price

Tied to the emerging research on distribution-level markets is a need to improve the pricing mechanism for distribution market operation [26]. The pricing mechanism should have the following attributes: 1) coordinate with the existing wholesale market, 2) incentivize DERs to be properly operated and developed, 3) reflect the cost and physical operating conditions of the distribution system, and 4) reward DERs for flexibility and grid service provided. These attributes will facilitate the integration of DERs into distribution networks.

The dominant pricing method at the transmission level is the locational marginal price (LMP) which has been widely implemented across the U.S. and some other electricity markets worldwide. In essence, the LMP reflects the marginal cost of energy supply, active transmission security constraints, and transmission losses at each bus in the system [27]. This is a good reference for the distribution network. Many studies have been done to extend the LMP to the distribution locational marginal price (DLMP) to capture the temporal and spatial value of electricity.

The DLMP or distribution-level nodal pricing was first proposed in [28], in which the DLMP was distinguished from the direct current optimal power flow (DCOPF) based pricing by considering network losses. The value of DGs was quantified by reducing line losses and loading. The DLMP developed in [29] included energy, congestion, and power loss components to drive controls in a distribution system that employed electronic devices. In [30][31], the distribution system operator (DSO) determined the DLMP based on generation offers and load bids by clearing the distribution market, in which the DLMP was discovered in a market environment. In [8], the DLMP was extended to include voltage components by considering the voltage constraints. The contribution of DERs in distribution system operation via voltage support and loss reduction was rewarded by the DLMP. Reference [32] estimated the intervals of the DLMP and the confidence levels with the consideration of renewable generation. Reference [37] presented a detailed analysis of three approaches toward understanding the DLMP because the interpretation of the DLMP is critical for policymaking in the growth of future distribution markets. These works show that DLMP has received broad attention and that it can maintain price attributes and reflect the operational conditions and DER values in a distribution system.

### C. Motivation of this Review and Visionary Paper

In summary, the distribution-level market and DLMP are promising directions for the process of distribution system deregulation. Therefore, it is necessary and beneficial to have a comprehensive and in-depth understanding of the state-of-the-art works for researchers to conduct further studies and applications. This review and visionary paper presents a comprehensive survey of ongoing distribution-level market studies, such as market formulation, model relaxations, DLMP characteristics, solution methods, and applications of DLMP. This paper also envisions future research directions and possible barriers and challenges related to DLMP implementation.

The rest of this paper is organized as follows. Section II presents the market participants, two typical distribution-level market architectures, and pioneering programs. Section III describes the market-clearing model, relaxation methods, DLMP formulation, comparison between LMP and DLMP, market properties of DLMP, and comparison of the DLMP with other electricity tariffs. Widely used market-clearing solution methods are summarized in Section IV. The DLMP's impacts on operation and planning for DERs and the DSO are analyzed in Section V. Section VI and Section VII discuss potential future research directions and possible challenges. Finally, Section VIII concludes the paper.

## II. DISTRIBUTION-LEVEL MARKET: PARTICIPANTS, ARCHITECTURES, AND ONGOING PILOT PROJECTS

An electricity market is a place where transactions can happen between power suppliers and consumers. The power suppliers are diversified in an active distribution system, including PVs, WTs, MTs, ESs, combined heat and power (CHP), etc. Consumers, the individuals or communities of multiple end-users, are becoming proactive.

This section first describes the market participants. Then, based on different transaction forms, the distribution-level market is classified into two categories: the pool-based market and the peer-to-peer (P2P) market. Their strengths and weaknesses are compared and summarized. Finally, emerging industry pilot projects for the distribution market are presented.

### A. Market Participants

Participants in the distribution-level market can generally be classified into the following categories: distribution market operators (DMOs) [38], distribution network operators (DNOs), electricity retailers [39], DGs, ESSs, microgrids (MGs), large consumers (LCs), and load aggregators (LAs).

It should be noted that the first three entities, DMOs, DNOs, and electricity retailers, typically stem from conventional DSOs which are the companies in the regulated distribution industry. Therefore, the DMOs, DNOs, and electricity retailers may reside in the same holding company even in a deregulated environment. As such, many papers in the literature use the term DSO even though it may refer to one or more entities among DMOs, DNOs, and electricity retailers. In this paper, the generic term DSO will also be used for simplicity, although the discussion below distinguishes between DMOs, DNOs, and electricity retailers based on their different functions in the distribution market.

*1) DMO:* As a profit-neutral entity, a DMO provides a trading platform that enables transparent energy transactions between electricity producers and consumers, clears the market, broadcasts price signals, determines market settlement, and so on, at the distribution level.

*2) DNO:* As the network operator, a DNO is responsible for system planning and operation, power outage restoration, network security management, etc. It may be the owner of distribution networks as well. Since both DMO and DNO are regulated, they may both reside within the same holding company as different departments. In some literature, such as [104], the DNO charges a tariff to the energy sellers and buyers for network usage.

*3) Electricity retailer:* As a retail service provider, an electricity retailer interacts with the external grid to maintain the power balance of the distribution-level market (e.g., purchasing electricity from the wholesale market when there is an electricity deficit and selling electricity to the wholesale market when there is extra generation in the local market). The electricity retailer can also provide ancillary services to the transmission system operator (TSO).

*4) DGs:* DGs are generators with a relatively large power capacity that satisfies market access conditions in a distribution system. The typical DGs include MTs, PVs, WTs, CHP, etc. These DG types can participate in the market directly by offering a certain amount of electricity at a specific price [40].

*5) ESSs*: As the penetration level of ESSs increases, they become an important and integral part of a distribution system. An ESS can be regarded as a prosumer due to its dual roles of producer and consumer. ESSs are assumed to have enough rated power and rated energy to participate in the distribution market. They can operate in an arbitrage manner or provide ancillary services to receive revenue [41].

*6) MGs*: MGs are viewed as a localized group of small DGs, small ESSs, and loads. An MG is also a prosumer in that it exports or imports electricity through the tie-line that connects to main distribution network. Therefore, an MG can both provide offers when it has extra power generation and bid when its self-production is insufficient.

*7) LCs*: LCs, such as manufacturing factories and commercial centers, usually have a large load demand and can participate in the market directly [42].

*8) LAs*: Individual consumers are usually not eligible to participate in the market directly due to the complex market rules, strict participation requirements [43], small adjustable capacities, and heavy calculation burden [44]. Typically, an LA is proposed to manage the flexibilities offered by multiple consumers and act as the intermediary between individual consumers and the DSO. By collecting consumers' consumption preferences and operation status, the LA bids in the distribution market on behalf of consumers and dispatches the purchasing power to contracted consumers. By doing so, the LA not only provides flexibility to the distribution system but also explores new business models for individual consumers to increase energy efficiency.

*B. Pool-based Market*

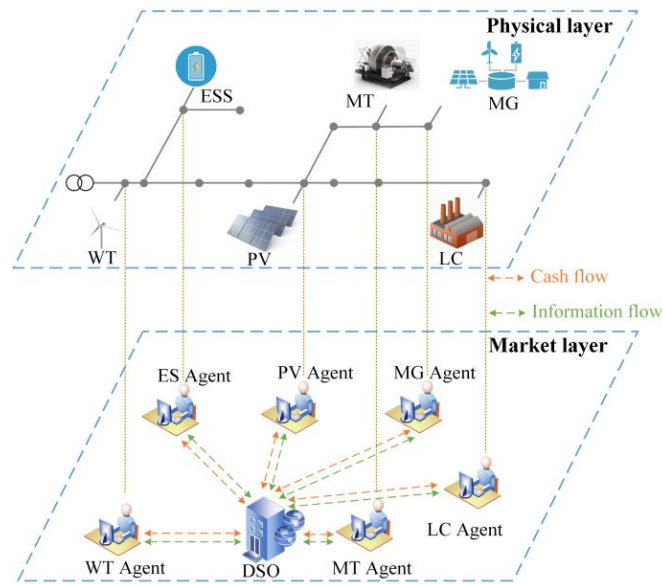

Fig. 2. Pool-based market architecture

The mission of the pool-based electricity market is to provide reliable electricity at the lowest cost to consumers [45]. Similar to the wholesale market, the transactions in a pool-based distribution market are centralized, hierarchical, and top-down. The DSO is responsible for the efficient market operation and coordination of various DERs. The DSO collects the bids and offers from all market participants, then clears the market in a centralized way, and provides incentives to the owners of DERs to further deploy these resources. The architecture of the pool-based market is illustrated in Fig. 2.

In References [8][30], the DSO cleared the market by solving the optimal power flow (OPF) problem with the objective of minimizing the system generation cost or maximizing social welfare. The optimal power quantity was broadcast to the participants, together with the DLMP that was derived from the dual variables. The influence of the uncertain DERs on the market and the DLMP was studied in [46][47]. In [34][35], the single-phase OPF formulation was extended to three phases, and the schedule of DERs in each phase was calculated.

*C. Peer-to-Peer Market*

The P2P market architecture is a less centralized, freer, and bottom-up electricity market where suppliers and consumers can autonomously transact electricity and other services. In this peer-centric architecture, each participant can negotiate, accept, or reject a trade according to their preferences, rationalities, and privacy considerations [48]. Demand and supply are finally matched by the specific distributed algorithms. A general P2P market architecture is illustrated in Fig. 3.

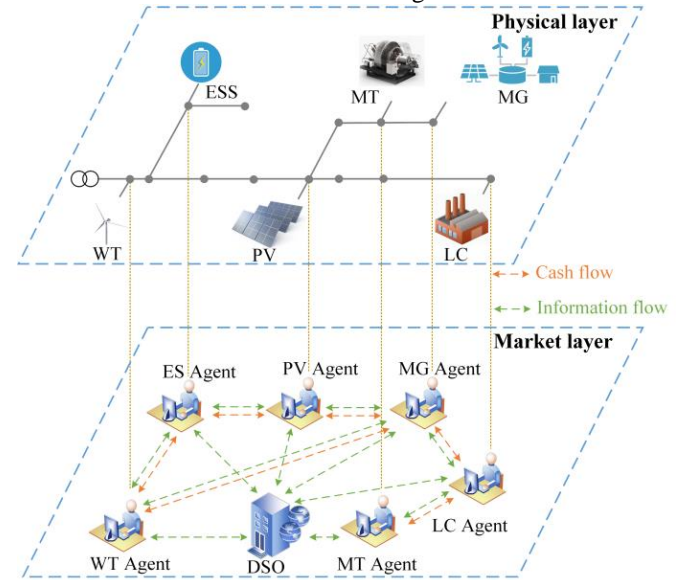

Fig. 3. P2P market architecture

Unlike the pool-based market, the design of a P2P market remains an open area and lacks consensus. Thus, this subsection intends to shed some useful light on market structure classification, physical network consideration, and DLMP deployment based on the existing literature. The market structure has been summarized and discussed in [49][50][51] and includes four types: 1) fully decentralized P2P markets where participants can directly negotiate with each other to complete trading; 2) coordinated P2P markets where trading is performed in a centralized manner in a community with a community manager; 3) community-based P2P markets where peers exchange limited information with the coordinator to preserve their privacy and autonomy; and 4) hybrid (or composite) P2P markets where the fully decentralized and the community-based P2P markets are combined.

In some other studies, network constraints have been considered, and the DSO is placed in the negotiation loop. In [48], a new P2P platform was conceptualized with the DSO as a delivery service provider. Peers first set up their trades by a peer matching algorithm, then the DSO checked the feasibility of the alternating current optimal power flow (ACOPF). If constraints were violated, the DSO updated the network usage price DLMP, and peers updated the cost function and redid the trade matching. This process continued until the ACOPF was feasible and all trades were unchanged. In this work, the DLMP promoted the P2P transactions that facilitated system operations and penalized those that were unfavorable from the perspective of the DSO. In [52], a two-stage P2P energy trading model was proposed. Trades were settled among peers in the first stage, then submitted to the utility for approval in the second stage. Certain trades were disapproved to ensure secure system operations. In [53], a coordinated P2P and ancillary service market was proposed, where the components of the DLMP were utilized as price signals for procuring ancillary services to ensure secure operation. In the auction-based P2P market proposed in [54], the DLMP was integrated into the payments of both prosumers and consumers to cover the power loss and voltage regulation costs.

Efficient algorithms are the key to enabling P2P transactions. Accordingly, various distributed algorithms and blockchain technologies have been proposed and developed to this end [55][56].

### D. Comparison Between Pool-based Market and P2P Market

Though the pool-based market and P2P market have different market architectures and market rules, they do share many similarities, including but not limited to: 1) boosting DER utilization; 2) increasing system resilience and reliability; 3) lowering consumers' payments; and 4) improving system operation efficiency. Their strengths and weaknesses are compared and summarized in TABLE I.

This paper is not going to claim that one market architecture is always more attractive than the other. Instead, we have provided a concise comparison study to give readers a basic understanding of these two markets.

### E. Distribution Market Pilot Projects

To take advantage of DER flexibility and receive economic benefits, governments, institutions, and investors have conducted multiple distribution-level market pilot projects. Based on the business model and research focus, three types of projects are reviewed here: the TSO-DSO coordination trial, the pool-based market, and the P2P market. Several representative projects are summarized in TABLE II.

The Transmission and Distribution Interface 2.0 [17] and SmartNet [18] mainly focus on TSO-DSO coordination schemes. In these schemes, the DSO manages flexible DERs in distribution to provide ancillary services to the TSO. Several key points regarding TSO-DSO coordination have been studied.

The Cornwall Local Energy Market [19], Piclo Flex [20], NODES [21], and EcoGrid 2.0 [22] are pool-based markets. The commonalities of these projects are that a local market has been created and the flexibilities are the main products traded in the market. Different price mechanisms, such as pay-as-bid and pay-as-clear, have been utilized in different projects.

EMPOWER [23] is a less centralized market, and it has studied approaches to exploring and developing ICT solutions to support local business models. The Brooklyn Microgrid [24] and P2P-SmarTest [25] are P2P markets. The Brooklyn Microgrid has created a platform where consumers and prosumers can trade self-produced energy locally in a P2P manner. P2P-SmarTest has simulated energy trading among interconnected MGs, as well as entities outside MGs.

## III. DLMP MODELS AND CHARACTERISTICS

Detailed reviews on the P2P market have been performed in [49][50][51]. To avoid repetition and to clearly present the DLMP derivation and explanation, here, a centralized pool-based market model is assumed and built.

As discussed in Section II, the distribution market involves various participants, and their bidding/offering activities and network physical constraints are essential for the secure and efficient operation of the market. This section summarizes the current, widely used centralized market models, relaxation methods, DLMP formulations, and DLMP characteristics compared to other tariffs. Note that although this market model mainly refers to the energy market, which is an essential part of electricity market operation in general, the ancillary service market can also be integrated into the current model as a co-optimization formulation.

### A. Market Clearing Model

A typical market-clearing model is shown below.

$$\min \sum_{i \in \Omega_G} f\left(P_i^G, \hat{Q}_i^G\right) \tag{1}$$

s.t.

$$\sum_{i \in \Omega_G} P_i^G - \sum_{i \in \Omega_N} P_i^D - P^{loss} = 0 : \lambda^p \tag{2}$$

$$\sum_{i \in \Omega_G} Q_i^G - \sum_{i \in \Omega_N} Q_i^D - Q^{loss} = 0 : \lambda^q \tag{3}$$

$$S_l^{\min} \leq S_l \leq S_l^{\max} : \omega_l^{s,\min}, \omega_l^{s,\max}, \forall l \in \Omega_L \tag{4}$$

$$V^{\min} \leq V_j \leq V^{\max} : \omega_j^{v,\min}, \omega_j^{v,\max}, \forall j \in \Omega_N \tag{5}$$

$$P_i^{G,\min} \leq P_i^G \leq P_i^{G,\max} : \omega_i^{p,\min}, \omega_i^{p,\max}, \forall i \in \Omega_G \tag{6}$$

$$Q_i^{G,\min} \leq Q_i^G \leq Q_i^{G,\max} : \omega_i^{q,\min}, \omega_i^{q,\max}, \forall i \in \Omega_G \tag{7}$$

$$-Q_i^G \leq \hat{Q}_i^G, Q_i^G \leq \hat{Q}_i^G : \kappa_i^-, \kappa_i^+, \forall i \in \Omega_G \tag{8}$$

where the objective function (1) is to minimize the total electricity generation cost including the electricity purchasing cost from the wholesale market and the generation cost of DGs; constraints (2) and (3) represent the active and reactive power balance constraints; (4) is the congestion constraint; (5) is the voltage constraint; (6) – (7) are the generators' active and reactive power output limits; and in (8), $\hat{Q}_{i,t}^G = \left| Q_{i,t}^G \right|$ since both absorbing and generating reactive power will induce cost [30].

The market-clearing model (1)-(8) provides a general and concise formulation. It can easily be extended to multiple time slots by adding time coupling constraints, and to more complicated formulations by considering different constraints and specific operating conditions. A detailed comparison of constraint considerations is summarized in TABLE III. It can be observed that in addition to constraints (2)-(8), other constraints have also been considered. For instance, in [46], the stochasticity of renewable DERs was internalized in the DLMP by applying conic duality to a chance-constrained ACOPF. In [47], the concept of an uncertainty DLMP was proposed to charge the uncertain resources and was seen as the marginal cost of dealing with the next unit increment of uncertainty in the distribution system. In [57], the uncertainties of renewable energies and load demand were priced in the LMP, which shed some light on pricing uncertainties in DLMP. The three-phase distribution system has been studied in [33][34][35][36], in which the single-phase DLMP was extended to three phases, and each phase had its own DLMP. The authors in [58] integrated the transformer loss of life in the relaxed branch flow model and added transformer degradation cost in the DLMP.

### B. Model Relaxation

Due to the high nonlinearity and nonconvexity of the ACOPF in distribution systems with high R/X ratios and power losses, it is generally hard to solve the clearing model, especially for a large system. To improve calculation efficiency as well as maintain accuracy, various methods have been proposed to approximate and relax the ACOPF model. The basic idea is to transform the nonlinear problem into a linear or convex problem. According to their mathematical properties, these approximation and relaxation methods can be categorized into two groups: linearization and convexification [59], which are summarized in TABLE IV.

In the linearization category, the nonlinear ACOPF model is approximated to a linear model via certain assumptions. A widely used model is the linearized *DistFlow* model [60], in which power loss and voltage angle are neglected. In [61], the quadratic calculation of branch power loss was linearized through the piecewise linear formulation. In [8][46], a polygonal approximation was used to linearize the quadratic congestion and power output constraints. Reference [59] proposed a modified *DistFlow* by using the ratios of active and reactive power to voltage magnitude as state variables, and then power flows and voltage magnitudes were written in a matrix form by introducing the path-branch incidence matrix. In [30][63], linearized power flow for distribution (LPF-D) and loss factors for the distribution (LF-D) were proposed to linearize the ACOPF, in which the active and reactive power were expressed as linear functions of voltage magnitudes and phase angles. In [64]-[66], the Taylor approximation was utilized to linearize the ACOPF model based on a pre-determined initial operating point. In addition to these model-based linearization approaches, data-driven power flow linearization approaches have been proposed by utilizing the power system measurement data [67][68]. In these approaches, with the determination of the constant terms and the regression parameter matrices, the active and reactive power injection were expressed as affine functions of voltage magnitudes and angles, or vice versa. Some studies have also focused on the linearization in the unbalanced system. In [62][69], the single-phase linearized *DistFlow* has been extended to an unbalanced three-phase model with the assumptions that voltages of three phases are similar and line losses are small. In [70], the three-phase power flow in the bus injection model was linearized for an active distribution network (ADN). In [35], LPF-D and LF-D were extended to three-phase systems.

In the convexification category, the ACOPF is usually convexified via two relaxation methods: semidefinite programming (SDP) [71]-[74] and second-order cone programming (SOCP) [75]-[77]. In [71], SDP was first proposed to relax the ACOPF. The authors in [72] proposed using SDP to solve the dual of an equivalent form of the ACOPF problem, and a necessary and sufficient condition was provided to ensure the zero-duality gap. In [73], SDP relaxation of the ACOPF was proposed for multiphase distribution networks with wye and delta connections. In [74], the unbalanced ACOPF problem was formulated as a moment relaxation-based SDP model, with system sparsity used to boost computational efficiency. In the SOCP category, the main idea was to relax the quadratic equality constraints into inequality constraints. The ACOPF was relaxed into a SOCP model for a radial network in the bus injection model in [75] and the branch flow model in [76]. The accuracy and exactness of the SOCP method have been proved in [76]. In case the SOCP relaxation is not exact, reference [77] has proposed a sequential SOCP approach to recover the actual optimal solution. References [78][79] have provided a detailed tutorial for the SDP and SOCP relaxation of the generic ACOPF problem. One can refer to [78][79] for more details regarding the formulation, derivation, and sufficient conditions to ensure the exactness of relaxations.

Additionally, some of the previously mentioned papers also present a numerical comparison regarding the accuracy of these approximation and relaxation methods. These papers are summarized here for readers convenience: [30][59][62][67][69][70][71][73][77].

### C. DLMP Formulation

Based on (1)-(8), the Lagrangian function of the clearing model can be written as follows.

$$
\begin{aligned}
L = &\sum_{i \in \Omega_G} f\left(P_i^G, Q_i^G\right) \\
&- \lambda^P \left(\sum_{i \in \Omega_G} P_i^G - \sum_{i \in \Omega_G} P_i^D - P^{loss}\right) - \lambda^q \left(\sum_{i \in \Omega_G} Q_i^G - \sum_{i \in \Omega_G} Q_i^D - Q^{loss}\right) \\
&- \sum_{l \in \Omega_L} \omega_l^{s,\min} \left(S_l - S_l^{\min}\right) - \sum_{l \in \Omega_L} \omega_l^{s,\max} \left(S_l^{\max} - S_l\right) \\
&- \sum_{j \in \Omega_N} \omega_j^{v,\min} \left(V_j - V^{\min}\right) - \sum_{j \in \Omega_N} \omega_j^{v,\max} \left(V^{\max} - V_j\right) \\
&- \sum_{i \in \Omega_G} \omega_i \cdot g_i\left(x\right)
\end{aligned}
\tag{9}
$$

where $g_i\left(x\right)$ represents the power output limits in (6)-(8).

TABLE I Comparison between pool-based market and P2P market

| Categories | Strengths | Weaknesses | Refs. |
|---|---|---|---|
| Pool-based market | 1) Market is straightforward and easy to implement.<br>2) System operating conditions (e.g., congestion, voltage) are easier to manage.<br>3) DERs investment is clear due to explicit market price signals. | 1) Aggregation technology is indispensable to aggregating multiple individual small-scale consumers.<br>2) Market power and strategic bidding of some participants exist.<br>3) DERs information (e.g., bidding price, quantity, and operating constraints) should be accessible to the DSO, so the protection of privacy is an imperative task. | [8][30] [34][35] [46][47] |
| P2P market | 1) It provides greater number of free choices among energy consumers and producers.<br>2) Information privacy of market participants can be preserved.<br>3) Consumers' preferences (e.g., cost, renewable, location) can be satisfied. | 1) Trading negotiation between peers is of high computational burden, and the scalability and convergence are challenges.<br>2) Behaviors of participants are less predictable to the DSO.<br>3) DERs investment becomes obscure due to the free and autonomous negotiations.<br>4) Network constraints should be addressed carefully. | [48][49] [53][54] [55][56] |

TABLE II Pilot projects of distribution level market

| Project name | Country/Region | Starting year | Business model | Price mechanism | Products to be traded | Project description |
|---|---|---|---|---|---|---|
| Transmission and Distribution Interface 2.0 [17] | United Kingdom | 2017 | TSO-DSO | NA | Flexibility services | Utilization of DERs flexibility to resolve the transmission voltage issues. |
| SmartNet [18] | European Union | 2016 | TSO-DSO | Pay-as-clear | Ancillary services | Comparison of different TSO-DSO schemes for acquiring ancillary services from DERs. |
| Cornwall Local Energy Market [19] | European Union | 2017 | Pool-based | Pay-as-bid and pay-as-clear | Flexibility services | A local-level market-based platform for flexibility trading among residential loads, CHP, ESSs, PVs, WTs, and DSO. |
| Piclo Flex [20] | United Kingdom | 2019 | Pool-based | Pay-as-bid | Flexibility services | A platform where DSO procures flexibility from DERs. |
| NODES [21] | Norway | 2018 | Pool-based | Pay-as-bid | Flexibility services | A market operator to test near-to-real-time flexibility procurement and minimize supplier imbalances. |
| EcoGrid 2.0 [22] | Denmark | 2016 | Pool-based | Pay-as-bid | Flexibility of aggregators | Development of a local flexibility market to demonstrate the grid-scale services provided by flexible residential loads. |
| EMPOWER [23] | European Union | 2015 | Less centralized | NA | Local energy | Exploration and development of an integrated ICT solution to support local markets and business models. |
| Brooklyn Microgrid [24] | United States | 2016 | P2P | Pay-as-clear, pay-as-bid, and fixed price | Local energy | Creation of a microgrid energy market where consumers and prosumers can trade self-produced energy locally. |
| P2P-SmarTest [25] | European Union | 2015 | P2P | Pay-as-clear | Energy among microgrids | Simulation of energy trading among interconnected MGs, as well as aggregators, DSOs, and the wholesale market. |

TABLE III Modeling details and comparisons

| Refs. | Power losses | Congestion | Voltage | Renewable energy uncertainty | Demand uncertainty | Phases | Transformer degradation | Active power | Reactive power | Reserve |
|---|---|---|---|---|---|---|---|---|---|---|
| [8] | ✓ | ✓ | ✓ | -- | -- | 1 | -- | ✓ | ✓ | -- |
| [30] | ✓ | -- | ✓ | -- | -- | 1 | -- | ✓ | ✓ | -- |
| [32] | ✓ | ✓ | ✓ | -- | -- | 1 | -- | ✓ | ✓ | -- |
| [81] | -- | ✓ | -- | -- | -- | 1 | -- | ✓ | -- | -- |
| [82] | -- | ✓ | -- | -- | -- | 1 | -- | ✓ | -- | -- |
| [83] | ✓ | ✓ | -- | -- | -- | 1 | -- | ✓ | -- | -- |
| [84] | ✓ | -- | ✓ | -- | -- | 1 | -- | ✓ | ✓ | -- |
| [85] | ✓ | -- | ✓ | -- | -- | 1 | -- | ✓ | ✓ | ✓ |
| [46] | ✓ | ✓ | ✓ | ✓ | -- | 1 | -- | ✓ | ✓ | ✓ |
| [47] | -- | -- | ✓ | ✓ | -- | 1 | -- | ✓ | ✓ | ✓ |
| [57] | -- | ✓ | -- | ✓ | ✓ | 1 | -- | ✓ | -- | ✓ |
| [58] | ✓ | ✓ | ✓ | ✓ | -- | 1 | ✓ | ✓ | ✓ | -- |
| [34] | ✓ | -- | ✓ | -- | -- | 3 | -- | ✓ | ✓ | -- |
| [35] | ✓ | -- | ✓ | -- | -- | 3 | -- | ✓ | ✓ | -- |
| [36] | ✓ | -- | -- | -- | -- | 3 | -- | ✓ | ✓ | -- |

TABLE IV Different relaxation methods

| Categories | Specific methods | Refs. |
|---|---|---|
| Linearization | DCOPF | [81][82] |
| | Linearized *Distflow* | [8][60][62][69] |
| | Polygonal approximation | [8][46] |
| | Modified *DistFlow* | [59] |
| | LPF-D, LF-D | [30][35][63][84] |
| | Taylor approximation | [64][65][66] |
| | Data-driven linearization | [67][68] |
| Convexification | SDP | [71][72][73][78][79] |
| | Moment relaxation-based SDP | [74] |
| | SOCP | [75][78][79][80][86] |
| | Sequential SOCP | [77] |

TABLE V Summary and comparison of solution methods

| Categories | Algorithms | Pros | Cons | Refs. |
|---|---|---|---|---|
| Centralized methods | Programming-based | It can find the global optimum if the problem is convex. | It has high computation burden in a large-scale problem. | [30][64][81][82][106] |
| | Metaheuristic | It can find a sub-optimal solution even if the problem is nonconvex and nonlinear. | It has high computation burden due to a large group of populations and a large number of iterations. | [80][107] |
| Distributed methods | ADMM | It can decompose a problem into sub-problems to improve computational efficiency. Data privacy is well protected. | The convergence rate of ADMM is not always satisfactory, and variants of ADMM are often needed for specific problems. | [109][110][111][112][113][114][115][116] |
| | ATC | It is suitable for hierarchical design optimization problems. | Implementation is complex, and it may not be able to handle a meshed network efficiently. | [117][118][119][120][121][122] |
| Decentralized methods | PMP | It has high computational efficiency due to in-parallel solution of sub-problems. | Information exchange among sub-problems is complex. | [123][124][125][126] |
| | APP | It makes decomposition possible in non-separable situations. | Information exchange among sub-problems is complex, similar to PMP. | [127][128][129][130] |

The active power DLMP and reactive power DLMP are the first-order partial derivatives of the Lagrangian function with respect to the active and reactive load demands, respectively.

$$\pi_i^p = \frac{\partial L}{\partial P_i^D} = \lambda^p + \lambda^p \cdot \frac{\partial P^{loss}}{\partial P_i^D} + \lambda^q \cdot \frac{\partial Q^{loss}}{\partial P_i^D}$$
$$+ \sum_{l \in \Omega_L} \left( \omega_i^{s,\max} - \omega_i^{s,\min} \right) \cdot \frac{\partial S_l}{\partial P_i^D} + \sum_{j \in \Omega_N} \left( \omega_j^{v,\max} - \omega_j^{v,\min} \right) \cdot \frac{\partial V_j}{\partial P_i^D}$$
$$(10)$$

$$\pi_i^q = \frac{\partial L}{\partial Q_i^D} = \lambda^q + \lambda^q \cdot \frac{\partial Q^{loss}}{\partial Q_i^D} + \lambda^p \cdot \frac{\partial P^{loss}}{\partial Q_i^D}$$
$$+ \sum_{l \in \Omega_L} \left( \omega_i^{s,\max} - \omega_i^{s,\min} \right) \cdot \frac{\partial S_l}{\partial Q_i^D} + \sum_{j \in \Omega_N} \left( \omega_j^{v,\max} - \omega_j^{v,\min} \right) \cdot \frac{\partial V_j}{\partial Q_i^D}$$
$$(11)$$

where $\pi_i^p$ and $\pi_i^q$ refer to the active DLMP and reactive DLMP at node $i$, respectively.

In (10), the active DLMP consists of four components: the marginal energy price, the marginal power loss price, the marginal congestion price, and the marginal voltage support price. The marginal price represents the additional cost of supplying the next MWh of electricity. Note that this additional cost can be decomposed into the subset of shadow costs associated with the binding constraints for that node. The marginal energy price is set by the marginal unit in the distribution system. The marginal congestion price reflects the marginal cost of congestion at a given node. Although congestion rarely happens in traditional radial distribution networks, it is possible in ADNs due to the high penetration of DERs or MGs. The marginal power loss price is set to reflect the power losses associated with power delivery which is not negligible because the power loss amount in a distribution system is usually higher than that in a transmission system. The voltage support price represents the cost of maintaining the voltage at an acceptable level, and it will be zero if the voltage constraints at a node are not binding. An in-depth understanding of the DLMP components can help stakeholders make better decisions, and is discussed in Section V.

There have been discussions about whether and how reactive power should be priced. In [8][30], the concept of a reactive DLMP which could act as a candidate choice to price the reactive power was proposed. In [87], the author stated that a reactive DLMP could be used as a tool to improve system stability and power quality. Thus, the reactive DLMP was kept for flexibility. If the reactive LMP is not considered in the transmission network, it is simply set to zero at the power supply point; if considered, such as the Q-LMP proposed in [88], (11) can be adopted directly to calculate the reactive DLMP.

It should be noted that the above formulation (10)-(11) is based on the typical market model (1)-(8). Additional price components such as marginal imbalance price [34] and transformer degradation price [58] can be included when other constraints are considered.

### D. Comparison Between LMP and DLMP

The LMP and DLMP share a lot of similarities, from their derivation and their physical meaning. For instance, they both have the combination of shadow prices of binding constraints, and both reflect the marginal cost at a node. However, they do have several differences which stem from the physical differences between the transmission and distribution networks. These can be summarized in the following four aspects.

*1) Congestion*: Line congestions occur more frequently in the transmission network due to the meshed structure and bulk electricity transactions. A feeder or lateral in a radial distribution network is usually designed to undertake the peak power flow, which makes it very rare for congestion to occur. However, in the future, if there is a higher penetration of DERs and MGs with possible network reconfigurations, power flow patterns may change with possible congestion issues similar to transmission systems.

*2) Power losses*: A distribution line usually has a higher R/X ratio and thus a higher power loss factor than a transmission line. Thus, the amount of power losses is not negligible at the distribution level while it is not unusual to omit losses at the transmission level when DCOPF is employed.

*3) Voltage*: The voltage drop is a more severe issue in the distribution system due to its radial topology with long feeders. Additional measures are necessary to maintain the system voltage in the acceptable range, and the corresponding cost should be reflected in the DLMP. In contrast, the wholesale LMP model may simply assume a flat voltage magnitude in the DCOPF model.

*4) Imbalance*: Imbalance is a unique feature of distribution systems, while transmission systems are generally considered balanced. Thus, a proper DLMP model should be extendable to incorporate multiple phases and include the cost of specific strategies to alleviate the imbalance level.

In addition to marginal energy price, the above discussion reveals the distinctions and the possible price components in LMP and DLMP. However, the number of price components in LMP or DLMP is not fixed. Depending on the application and the studied distribution network, one or more specific pricing components may be ignored.

*E. Market Properties of DLMP*

DLMP, extended from LMP, shares many common market properties, including Pareto efficiency, revenue adequacy, generation cost recovery, and incentive compatibility [89][90]. When the market objective is to maximize social welfare, the market equilibrium is achieved with Pareto efficiency because no Pareto improvement can be made [43]. Revenue adequacy indicates that the revenue of the market operator can cover the expenses of market operation. Generation cost recovery means that the electricity producer's profit is non-negative when DLMP is used for settlement. Incentive compatibility implies that when each participant pursues its best profit by following the market rules, all their behaviors promote the achievement of group goals.

It should also be noted that the above analysis assumes a perfectly competitive market. The actual market environment could be imperfect, and participants with market power can cause an inefficient market. However, despite the possibility of inefficiency, this assumption still works well in many markets due to the "invisible hand". Additionally, the above analysis also sheds light on the policies that aim to improve the market environment.

*F. Comparison Between Other Electricity Tariffs and DLMP*

Suitable pricing schemes are essential in distribution markets, and many other tariffs have also been studied in the literature. Based on the usage and design purposes, they can be classified into three categories: 1) DR-based prices, 2) P2P incentivizing prices, and 3) Nash bargaining (NB) -based prices.

*1) DR-based prices*: This category includes price-based or incentive-based DR programs. The former mainly consists of time of use (TOU), critical peak pricing (CPP), and real-time pricing (RTP), while the latter includes direct load control, interruptible loads, and peak time rebate (PTR) [91]-[96]. These economic signals via either financial incentives (i.e., coupons or rebates) or dynamic prices are proposed mainly to incentivize consumers to change their power consumption to relieve the operating stress of the power system.

The main difference between these electricity tariffs and the DLMP lies in how the electricity rate is generated. Fundamentally, DR-based prices or tariffs are exogenously generated by an electricity retailer or a DSO, possibly via a LA, conventionally in the regulated industry as a means to increase demand flexibility. In contrast, the DLMP is endogenously generated in distribution markets' bidding and offering process. It is a market-determined price reflecting the marginal or incremental cost of electricity services provided (e.g., power injection or withdrawal at a specific moment and location). In this sense, they arguably may not be very comparable to each other.

Technically, since power losses and voltage constraints are considered in the market clearing, the DLMP varies in both time and location. Also, the DLMP is composed of the shadow prices of the operating constraints, thus it exactly reflects the physical operating conditions of the system. In contrast, the operational constraints and load quantity variations barely affect the DR-based prices which are pre-determined. These DR programs are simple and straightforward to implement, and they do offer some flexibility to end consumers. However, their temporal granularity is usually coarse (e.g., peak and off-peak price for TOU, peak prices for CPP), and their spatial granularity is not reflective of network topologies, while the DLMP can increase the spatiotemporal granularity of electricity tariffs.

Nevertheless, they are not necessarily mutually exclusive and can be coupled or co-exist in the same distribution market. For instance, an LA and its DR participants can adopt a DR-based price scheme. In this way, the DLMP can work as an indicator for the LA to set up proper DR prices to recover the operation cost and the payment of purchasing electricity from the distribution market. Meanwhile, the revenue obtained from DR prices can also provide information for the LA bidding prices when the LA participates in the market. For example, in [97], TOU was used to charge power consumers, and the DLMP was generated as the power trading price between DGs and electricity retailers.

*2) P2P incentivizing prices*: These pricing schemes aim at incentivizing trading energy internally among peers in a P2P community. Three examples are: the mid-market rate (MMR) scheme [98][99], the bill sharing (BS) scheme [100][101], and the single linear programming (LP) pricing scheme [102].

The MMR and single-LP scheme can decrease the local electricity buying price and increase the local electricity selling price for DERs in a P2P community, such that the P2P trading is more economical than DERs trade directly with the external grid. Thus, prosumers will pay less and receive more revenue when trading in the community market. These price mechanisms attract prosumers to trade internally within the community rather than directly interacting with the external grid. The BS scheme reduces both the local buying price and selling price. Thus, consumers' payments are reduced, but producers will experience a revenue deficit. As a result, DGs may not be motivated to participate in P2P energy sharing, especially those who own a significant local generation. In addition, the BS pricing scheme offers a flat rate, and its load shifting effect is limited.

Compared to the DLMP, these three pricing mechanisms currently only reflect the power balance constraint in a P2P market. Further study is needed to integrate other physical operating constraints (e.g., power losses, voltages, congestions) in the model.

*3) NB-based prices*: Some studies have used game theory to model energy trading. For example, NB was proposed as an incentive mechanism to encourage interactions and benefit-sharing among MGs [103]. In [104], the network usage price was derived in an NB-based market to cover the cost of the network operator. It is an alternative to the DLMP working as the network usage price.

Based on the above descriptions and comparisons, price schemes in different market structures and designs may vary. For example, the DLMP can work as a general pricing method and be used in the settlement of power producers and consumers and the network usage charge, while P2P incentivizing prices, such as MMR and single-LP, can be employed for internal trading within a community or MG. NB-based prices can be used to achieve fair benefit allocations.

In summary, while the DLMP is the dominant pricing approach in the literature for modern distribution markets, other pricing schemes can be utilized for different market designs or co-exist with the DLMP.

## IV. MARKET CLEARING SOLUTION METHODS FOR DLMP

The DLMP is usually recovered from the dual variables when the market-clearing model is solved. Thus, solving the OPF problem of the distribution system is the prerequisite of the DLMP. This section reviews the solution methods of the market-clearing model. In the literature, these methods can be categorized into three groups: centralized methods, distributed methods, and decentralized methods. Their features are summarized and compared in TABLE V.

### A. Centralized Methods

Centralized optimization is defined as a coordinated policy for solving a problem that considers the system under study as a whole. In power systems operation, it means the control center collects all required information and performs centralized decisions [105]. The clearing model from Section III can be classified into different optimization models based on mathematical properties, such as LP, quadratic programming (QP), mixed-integer linear programming (MILP), SOCP, nonlinear programming (NLP), etc. Various centralized solution methods have been proposed and can be generally classified into either mathematical programming-based algorithms or metaheuristic algorithms.

#### 1) Mathematical Programming-based Algorithms

Programming-based algorithms, like the simplex method, interior point method, and branch and bound method, can find the optimal solution for typical OPF problems. Various commercial solvers that have integrated these algorithms can also solve the clearing model effectively and efficiently, such as CPLEX, GUROBI, MINOS, and IPOPT. The main advantage of a programming-based algorithm is that the global optimum can be guaranteed if the problem is convex. The main drawback is that the computation time increases significantly with the scale of the problem. In [30][81], the market model was linearized to form an LP problem that is solved effectively. In [82], the system operation cost and electric vehicle (EV) aggregator operation cost were modeled as quadratic functions, thus the market model belongs to QP. The ACOPF was convexified as a SOCP relaxation method in [64] and solved by GAMS. Some NLP methods have also been proposed. In [106], the trust-region method was utilized to solve the ACOPF, in which market equilibrium was obtained and the DLMP was recovered.

#### 2) Metaheuristic Algorithms

Metaheuristic algorithms, such as the genetic algorithm (GA), particle swarm optimization (PSO), and teaching-learning-based optimization (TLBO), are powerful in the sense that they can be applied to most optimization problems. However, it should be noted that these algorithms do not guarantee that the final solution is the optimal one. The pros are that they can find a sub-optimal solution for a nonlinear and nonconvex problem that mathematical programming-based algorithms cannot solve directly. The cons are that these algorithms are time-consuming because a large group of populations and many iterations are needed. Reference [80] proposed a methodology to achieve the optimal network reconfiguration with the application of the DLMP in distribution networks by utilizing GA. In [107], a modified PSO approach was proposed to obtain the optimal locations for the EV charging stations in a two-area distribution system.

### B. Distributed Methods

As the number of participants in the distribution market increases, it is challenging for centralized optimization methods to solve large-scale problems due to the computational burden. Also, data privacy is another concern for participants. Therefore, various distributed optimization methods have been proposed [108]. The basic idea of distributed optimization is to decompose the original model into a set of smaller sub-problems and solve each sub-problem independently till convergence, usually facilitated by a coordinator to exchange the information of coupling variables among sub-problems. As such, the computational burden is alleviated. Two commonly used distributed algorithms are reviewed: the alternating direction method of multipliers (ADMM) and the analytical target cascading (ATC).

### 1) Alternating direction method of multipliers (ADMM)

The ADMM is an augmented Lagrangian relaxation-based algorithm that can solve arbitrary-scale optimization problems and support distributed computation [109]. It utilizes the dual decomposition method to break a large-scale problem into smaller ones and the augmented multipliers method to deal with the convergence issue. The ADMM algorithm needs a central coordinator to update the dual variables. Overall, the ADMM is a robust algorithm that does not require strong assumptions and has good data privacy protection.

The ADMM has been applied frequently to power systems problems. Reference [110] utilized the ADMM algorithm to clear a market with high penetration of residential loads, in which residential LAs communicated with the DSO with limited information and thus protected residents' privacy. In [111], a consensus-based ADMM approach was proposed to solve the DCOPF problem with DR. Three types of ADMM-based distributed algorithms were performed, and different convergence performance and communication requirements were discussed. In [112], the ADMM algorithm was utilized to clear the market to protect the privacy of EVs and renewable DGs. Reference [113] proposed a consensus-based Jacobian ADMM algorithm to maximize grid welfare in a distributed and parallel manner, in which an RTP price scheme was derived to facilitate automated DR. In [114], the ADMM was utilized to clear a distribution-level flexibility market in which aggregator payback condition is considered. In [115], a multi-block ADMM algorithm was proposed to solve the market-based multi-period ACOPF, and a distributed Douglas-Rarchford Splitting method was utilized to ensure convergence. In [116], a consensus-ADMM structured trust-region algorithm was proposed to solve the nonlinear and nonconvex ACOPF for multiple regions.

### 2) Analytical target cascading (ATC)

The ATC algorithm is a system design approach enabling top-level design targets to be cascaded down to the lowest level of the modeling hierarchy [117]. In the ATC method, the entire system is partitioned into a set of subsystems that are hierarchically connected [118]. No coupling variables exist among subsystems at the same level; instead, coupling variables only exist between systems at adjacent levels.

In [119], the ATC theory was introduced in the economic dispatch of an ADN. The ATC algorithm decoupled the dispatching of distribution networks and multi-MGs by considering them as different stakeholders that could independently optimize their operation and economic benefits. In [120], the ATC method was adopted to achieve win-win cooperation between the ADN and virtual MGs in a coupling bi-level dispatching model. In [121], an accelerated robust ATC method was proposed to solve the OPF problem. A function was designed to determine a balancing coefficient to avoid premature convergence or divergence and reduce the number of iterations. In [122], EV charging stations and the DSO were formulated as a bi-level model. Then the ATC algorithm was proposed to decouple the bi-level model with the tie-line power as the only exchanged information.

## C. Decentralized Methods

Similar to distributed methods, decentralized methods also decompose a large-scale problem into smaller ones. The main difference is that distributed methods still require a coordinator that exchanges limited information of coupling variables among sub-problems, while decentralized methods do not involve a central coordinator, and sub-problems only exchange information with their neighbors [118]. Here, two decentralized algorithms are reviewed: the proximal message passing (PMP) and the auxiliary problem principle (APP).

### 1) Proximal message passing (PMP)

As stated in [118], the PMP algorithm is a special application of the ADMM. However, unlike the sequential sub-problem solving of the ADMM, the PMP algorithm is fully decentralized in which sub-problems are solved in parallel. Since the PMP algorithm inherits from the ADMM, all convergence properties that hold for the ADMM also hold for the PMP algorithm [123]. In [124], PMP algorithms with different penalty factors were tested on a distribution market-clearing problem. Each device optimized its sub-problem parallel to each other and exchanged information with neighbors at the end of each iteration. Reference [125] addressed the convergence issue of the PMP algorithm caused by the binding voltage constraints and discovered the DLMP. In [126], the APP algorithm and PMP algorithm were combined to solve a multiple-period OPF problem. The APP algorithm first distributed the computation to several dispatch periods, then the PMP algorithm solved each period across different power devices.

### 2) Auxiliary problem principle (APP)

In [127], the APP algorithm was proposed to decompose a problem into auxiliary problems with shared variables. Each auxiliary problem interchanged information with its neighbors via shared variables and no central coordinator was required. The APP algorithm has been applied in power system optimization problems. In [128], the whole power system was decomposed into multiple areas, and fictitious nodes were added to decouple the neighboring areas. Each area was an independent OPF problem, so it interchanged with neighbors at the end of each iteration until the convergence criteria were satisfied. In [129], the APP algorithm was utilized to solve a DCOPF problem of multiple interconnected systems. Each subsystem was solved independently with limited information exchange, avoiding sharing the local generation cost information. Though the APP is utilized in meshed networks in [128][129], similar approaches can be extended to radial networks. Reference [130] partitioned the ADN into different voltage control zones, then proposed an APP-based algorithm to solve each zone with limited data exchange.

## V. APPLICATIONS OF DLMP

In a market environment, the DLMP is an effective economic signal that can both reflect the marginal operating cost of a distribution system and provide continuous market signals for the operation and future investment of DERs. It has the possibility to change the conventional operation and planning pattern of the distribution system. In this section, the impacts of the DLMP on system flexibility, operation, planning, and reliability are reviewed from different perspectives.

### A. Boost Flexibility

Electric flexibility refers to adjusting electricity production or consumption in response to system variability or activation signals. In general, it has five attributes: (a) direction (up or down), (b) power capacity, (c) starting service time, (d) service duration, and (e) location [131]. These attributes can be utilized properly to provide specific services to the system.

In a distribution system, DERs are the primary sources of flexibility, and they can be divided into the supply and demand sides. Some power suppliers, such as WTs, fuel cells (FCs), and ESSs, are dispatchable, and they can work in coordination with each other to foster system flexibility [132]. On the other hand, consumers provide flexibility from the load side. This load-side flexibility is released by adjusting consumers' consumption behaviors via incentives or price signals [133].

In a regulated system with a flat tariff, DERs usually behave as price takers. They passively receive the electricity price and do not have much motivation to improve their consumption or generation schedules. DERs flexibility is obstructed if there is no proper pricing policy. Other tariffs, such as TOU, PTR, CPP, and RTP, have been utilized to incentivize DERs to release some flexibility. However, the potential of DER flexibility has not been fully explored and can be further released by suitable market mechanisms.

With the continuous improvement study of the DLMP and its spatial-temporal and operating condition reflective characteristics, the deployment of the DLMP can incentivize DERs to unlock their flexibilities for better benefits. How the boosted flexibility will influence DERs operation, planning, and system reliability are analyzed as follows.

### B. Incentivize DER Operations

With the DLMP in place, energy consumers and producers can make decisions based on incentives.

On the demand side, consumers wish to reduce their electricity procurement costs. This can be done by optimizing the time-shiftable loads under the condition of dynamic electricity pricing [134]. Consumers can shift electricity consumption from peak price hours to low price hours. Time-shiftable loads include charging of EVs, washing machines, dryers, etc. Among all residential loads, thermostatically controlled loads (TCLs) (e.g., heating, ventilation, and air conditioning (HVAC) systems and electric water heater) and EVs account for a large share of the electricity consumption and have similar energy storage characteristics [4][135]. These characteristics render them the ideal residential DR candidates that can respond to varying price signals. In [95][133][136][137], precooling or preheating for TCLs has been proposed to reduce electricity bills by avoiding consuming electricity during peak price hours. A bi-level model was proposed in [64] to optimize EV charging schedules.

On the supply side, DGs can provide offers at a marginal cost to maximize their profits [28]. PVs and WTs usually have low marginal costs, so they can easily win the bid, which will promote renewable penetration. Other DGs, such as MTs, FCs, CHP, and geothermal generators, are usually dispatchable but have a high marginal cost. Therefore, they can bid in the energy market to provide energy at peak hours to earn revenue since the DLMP in these hours is usually high, or they can participate in the ancillary service market to maintain the stable operation of the distribution system [138].

For prosumers in the system, such as ESSs and MGs, consumption/generation schedules can also be affected by the DLMP. Beyond their ability to restrain the uncertainties of renewables, prosumers can also operate in an arbitrage manner by utilizing the temporal difference of the DLMP. That is, prosumers buy electricity at low DLMP hours and sell electricity at high DLMP hours. In [47][139], bi-level models were proposed for MGs and the DSO. As the price taker, the optimal response of MGs under the DLMP has been studied. In [38], regarding the DLMP as the bidding price of MGs and the clearing price of the DSO, a bi-level model was built and a bidding strategy was proposed for MGs to minimize their operation cost and maintain the secure operation of the distribution system.

### C. Guide DER Planning

In general, DER planning focuses on minimizing both investment and long-term operating costs as well as satisfying systems operational requirements, such as meeting the demand growth, providing sufficient reserve, and reducing power losses, typically from the perspective of a DSO or a utility [140]-[144]. However, with the liberalization of distribution systems, the new pricing mechanism can provide stakeholders with incentive market signals. Given the DLMP's spatial and temporal properties, private investors will be motivated to build DERs (e.g., renewable DGs, ESSs) at ideal sites with optimal sizes to maximize their profits. They become the primary decision-makers during the planning phase.

Some studies have investigated this new planning scenario. In [145], a game-based model for the multiperiod planning of MGs in a competitive market was presented. The best locations were first determined based on the weighted sum of the loss sensitivity factor and voltage sensitivity factor, and then a bi-level model was used to find the best installation time and DER types. In a competitive market with specified candidate sites, Reference [146] selected the best sizes for renewable DGs and ESSs. In [147], a robust adaptive model for DER planning was proposed in which the 8760-hour operational conditions in each planning year were reduced to a solvable number. In [148], an exhaustive search approach was proposed to identify the most suitable WTs allocation and installation priority with fixed sizes. In [149], the DLMP was utilized to guide new line constructions and DER planning to maximize asset owners' profit. Reference [150] proposed that the DLMP can provide effective market signals for future unit investment in a deregulated distribution system. A two-stage stochastic bi-level programming model was formulated for investors to best site and size battery energy storage systems.

It should be noted that some of these studies [145]-[148] employed price signals to direct the DER planning, though none used a full-fledged DLMP algorithm. Nevertheless, they do supply some insights for this transition.

### D. Improve System Reliability

From the viewpoint of the DSO, the introduction of the DLMP is a promising method to improve the system operating condition. With proper incentive regulations, DERs have changed their consumption/generation behaviors as discussed in previous subsections. These changes in turn greatly benefit the distribution system, with the benefits including but not all limited to load shifting, peak shaving, reactive power support, renewable energy integration, carbon emission reduction, investment deferral, congestion, and voltage management. The system stability and resilience are improved, which is aligned with the incentive compatibility.

In [151], the DLMP was used to reduce a network's power losses and emissions. In [64][81][82], the temporal characteristic of the DLMP was utilized to optimize the charging schedule of EVs to alleviate the congestion issue. In [83][133][152], the DLMP was proposed to guide household DR to prevent congestion. EVs and PVs were studied to prevent congestion in [153]. The DLMP was raised to a high value in these studies when congestion occurred, incentivizing EV and DR owners to lower their demands. A similar principle has been applied to improve the voltage profile. In [84], it was demonstrated that the DLMP could also be very high when voltage constraints were binding. Thus, flexible HVAC and EV demands were shifted to low DLMP periods to alleviate the voltage violation. At the same time, DGs could increase their power production at these hours to improve their revenue. In the summers of 2020 and 2021, the California Independent System Operator (CAISO) advised consumers to turn off unnecessary appliances and defer the use of major appliances to alleviate the shortage of power supply, such as pre-cooling buildings before the afternoon [154]. In a deregulated system, consumers will likely take the initiative to perform these price-responsive activities according to the time-varying DLMP.

## VI. FUTURE DIRECTIONS

Based on the above analysis and review of the distribution-level market and DLMP, the following topics and directions related to DLMP are believed to be worth investigating by the academic and industrial communities in the future.

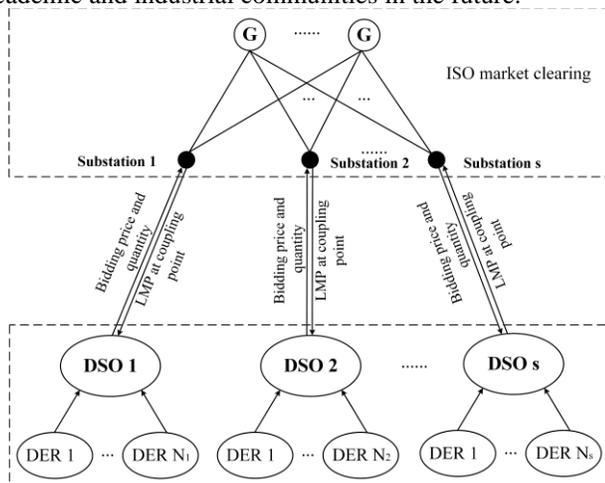

Fig. 4. Framework of DLMP considering coordination between transmission and distribution markets

### A. Coordination Between Transmission-level Market and Distribution-level Market

With the increasing penetration of DERs, the role of traditional distribution networks as load-serving entities in wholesale electricity markets now evolves towards ADNs, which can proactively participate in wholesale markets. The proliferation of DERs enhances the interactions between distribution and transmission-level markets and needs coordination between the two markets to facilitate market clearing and pricing.

A framework for a hierarchical market structure shown in Fig. 4 could capture the interactions between transmission and distribution markets [155]. In addition, the FERC Order 2222 [40] allows DERs and DER aggregators to participate in the wholesale market directly. Thus, it is worthwhile to explore and investigate effective pricing mechanisms, new market products, and participation models that could enable DERs to maximize their value and services to both distribution and transmission systems in a coordinated manner.

### B. Integration with P2P Energy Trading

P2P energy trading is emerging in the distribution system as the number of prosumers and DERs grows. P2P trading may happen among prosumers in a local community/microgrid or among aggregators in the distribution system. P2P has many advantages as stated in TABLE I, but it may not work as a standalone solution because the trading agents are myopic to network conditions and thus P2P trading may violate system security constraints. Also, not all customers may participate in P2P trading. Therefore, the mutual impact between P2P trading and the distribution market and DLMP needs to be fundamentally examined and understood. For example, when the network constraints are binding, in what priority should the participants in P2P trading or non-P2P DERs/prosumers be dispatched? How much of the system regulation cost should the participants in P2P trading bear? The DLMP could be designed to accommodate or align with P2P trading in distribution systems that retain the properties of P2P trading, such as preserving privacy and trading preferences, and meanwhile ensure system operational security and price fairness to all customers.

### C. Profit-oriented Investment

As discussed in Section V.C, pricing mechanisms can be effective market signals for infrastructure or DER investment. Due to cost-reflective and spatiotemporal characteristics, the DLMP can provide continuous and explicit incentives for system planners, stakeholders, and investors who intend to install DGs, MGs, or ESSs, or otherwise upgrade system infrastructure. The traditional planning problem in a regulated distribution system is transitioned to a profit-oriented one to make more profits and recover the investment costs. Pioneering works [145]-[150] have tried to use price signals to incentivize system planning which still needs further study in a more competitive environment. Especially with the employment of DLMPs as incentive signals and the emerging various trading forms in modern distribution systems, broad and in-depth studies of profit-oriented DER planning and investment are necessary and attractive.

*D. Pricing Uncertainty*

With more DERs integrated into the distribution network, it is essential to consider the uncertainty in the distribution market. The current deterministic optimization-based model might not be able to handle the high dimensional uncertainties of DERs in the future grid. Also, most works focus on addressing uncertainties to ensure system security but increase the operating cost. All participants generally share this additional cost, which is not truly fair. Based on the rule "who causes the problem is the one who should be responsible," uncertainty sources should undertake this part of the cost. In [46][47][57], uncertainty has been priced or internalized into the DLMP or LMP. These pioneering works provide helpful insights for uncertainty pricing. Thus, the definition of uncertainty prices and how to utilize uncertainty prices to quantify the impacts of DER uncertainty on the local voltage violations, line overloading, and DER service deliverability are topics deserving further study.

VII. POSSIBLE BARRIERS AND CHALLENGES OF IMPLEMENTING DLMP

Despite the advantages mentioned above and the promising future of the distribution-level market and DLMP, there are still several important potential barriers and challenges to implementing the DLMP in practice.

*A. Aggregation Technology*

Within a high DER penetration system, the DSO does not have the same level of visibility, control, and situational awareness of DERs on its system as the ISO does with transmission-connected generators [156]. Thus, it is necessary to develop new communication and control technologies that can aggregate a huge number of small DERs into a large virtual aggregator to participate in distribution markets or provide grid services. Furthermore, infrastructure is indispensable for a fully competitive distribution market. Therefore, before such technologies and infrastructure are mature and can be widely applied, it may be a little early to extensively implement the distribution market and the DLMP.

*B. Variability and Acceptance of DLMP*

The DLMP is influenced by the substation LMP, nodal load demands, and power output of DGs [84]. These factors vary widely with time, thus, the DLMP also varies significantly within a day, days, weeks, months, seasons, etc., making a general daily DLMP pattern hard to obtain. In addition, the distribution system operates in a standard network configuration in normal conditions, but it reconfigures the network in emergency events, such as heavy loads and power outages. Network reconfiguration may further change the DLMP pattern. These characteristics complicate the electricity price both temporally and spatially. Compared to the flat electricity rate and TOU that are fixed over a long time, the DLMP may be more complicated for power consumers to navigate. Thus, the price mechanisms should be carefully designed to consider the actual reactions of the consumers, and a LA should play an essential role in bridging the DSO and consumers. Studies in consumer psychology may be conducted, and a survey on the acceptance of the DLMP may be performed [157].

*C. Investigation of Policy to Ensure Fairness and Efficiency of Distribution Markets*

Due to the typical radial structure of distribution networks, the power consumption in the upstream will significantly influence that in the downstream. For instance, if consumers close to the substation (i.e., upstream) consume a large amount of energy, the amount of energy that distant consumers can obtain is limited due to line capacity and voltage limits. Meanwhile, distant consumers need to pay more if the operating limits bind. A few studies have proposed mechanisms to mitigate such an issue. In [158], hedging rights were utilized to mitigate the undesirable effects of increased DLMP to aggregators. In [159], Jain's fairness index was applied in the DLMP mechanism to achieve fair cost allocation. However, these methods do not fully resolve the fairness issue. It is still an important topic which needs further investigation.

*D. Market Power in Distribution Markets*

Currently, most studies in the existing literature have been carried out with the assumption of a perfectly competitive market. The distribution-level market could be imperfect in the real world because the number of large-scale participants is possibly limited. Further, different from the wholesale market, the unique characteristics of the distribution market, such as voltage and imbalance, can impact the DLMP considerably. These factors enable participants to exercise market power for better profit, such as increasing the clearing price. Thus, regulatory authorities need to propose measures to prevent or mitigate the abuse of market power. So far, these issues are rarely studied at the distribution level. Therefore, the market power-related topics, such as market power assessment, market power exercise, and mitigation measures, must be investigated to achieve an efficient distribution level market.

VIII. CONCLUSIONS

DLMP-based distribution-level electricity markets provide an effective solution to managing large amounts of DERs and MGs. Many pioneering works have been done on this topic in recent years. Hence, a summary of the state-of-the-art research on this topic is necessary to promote future studies in this area. This paper first outlined and reviewed the current progress toward the development of the distribution-level market in both academic and industrial communities. Second, a market-clearing model was established with specific distribution-related operation constraints and relaxations. Third, the DLMP was expressed explicitly, and its features with respect to other electricity tariffs were discussed. Then, this paper reviewed the state-of-the-art solution methods to solve the market clearing model. Further, various DLMP-related applications in distribution system operation and planning were discussed. Finally, our visions on several paths for future research as well as possible barriers and challenges were presented.

REFERENCES

[1]  K. Horowitz *et al.*, "An overview of distributed energy resource (DER) interconnection: current practices and emerging solutions," National

Renewable Energy Laboratory, Denver, CO, USA, Tech. Rep. NREL/TP-6A20-72102, April 2019.

[2] J. J. Cook, K. Ardani, E. O'Shaughnessy, B. Smith, and R. Margolis, "Expanding PV value: lessons learned from utility-led distributed energy resource aggregation in the United States," National Renewable Energy Laboratory, Golden, CO, USA, Tech. Rep. NREL/TP-6A20-71984, Nov. 2018.

[3] R. R. Mohassel *et al.*, "A survey on advanced metering infrastructure," *Int. J. Elect. Power Energy Syst.*, vol. 63, pp. 473-484, 2014.

[4] Q. Shi *et al.*, "Estimating the profile of incentive-based demand response (IBDR) by integrating technical models and social-behavioral factors," *IEEE Trans. Smart Grid*, vol. 11, no. 1, pp. 171-183, Jan. 2020.

[5] J. Hao, D. W. Gao, and J. J. Zhang, "Reinforcement learning for building energy optimization through controlling of central HVAC system," *IEEE Open Access Journal of Power and Energy*, vol. 7, pp. 320-328, 2020.

[6] X. Kou *et al.*, "Model-based and data-driven HVAC control strategies for residential demand response," *IEEE Open Access Journal of Power and Energy*, vol. 8, pp. 186-197, 2021.

[7] K. E. Lonergan and G. Sansavini, "Business structure of electricity distribution system operator and effect on solar photovoltaic uptake: An empirical case study for Switzerland," *Energy Policy*, vol. 160, pp. 112683, 2022.

[8] L. Bai, J. Wang, F. Li, *et al.*, "Distribution locational marginal pricing (DLMP) for congestion management and voltage support," *IEEE Trans. Power Syst.*, vol. 33, no. 4, pp. 4061-4073, July 2018.

[9] S. Yousefi, M. P. Moghaddam, and V. J. Majd, "Optimal real time pricing in an agent-based retail market using a comprehensive demand response model," *Energy*, vol. 36, no. 9, pp. 5716-5727, 2011.

[10] P. O. Rosell *et al.*, "Optimization problem for meeting distribution system operator requests in local flexibility markets with distributed energy resources," *Appl. Energy*, vol. 210, pp. 881-895, 2018.

[11] X. Jin, Q. Wu, and H. Jia, "Local flexibility markets: Literature review on concepts, models and clearing methods," *Appl. Energy*, vol. 261, pp. 1-35, 2020.

[12] C. A. Correa-Florez, A. Michiorri, and G. Kariniotakis, "Optimal participation of residential aggregators in energy and local flexibility markets," *IEEE Trans. Smart Grid*, vol. 11, no. 2, pp. 1644-1656, Mar. 2020.

[13] P. Goncalves Da Silva, D. Ilić, and S. Karnouskos, "The impact of smart grid prosumer grouping on forecasting accuracy and its benefits for local electricity market trading," *IEEE Trans. Smart Grid*, vol. 5, no. 1, pp. 402-410, Jan. 2014.

[14] Y. Parag and B. K. Sovacool, "Electricity market design for the prosumer era," *Nature energy*, vol. 1, no. 4, pp. 1-6, April 2016.

[15] F. Lezama, J. Soares, P. Hernandez-Leal, M. Kaisers, T. Pinto, and Z. Vale, "Local energy markets: Paving the path toward fully transactive energy systems," *IEEE Trans. Power Syst.*, vol. 34, no. 5, pp. 4081-4088, Sept. 2019.

[16] S. Chen and C. C. Liu, "From demand response to transactive energy: state of the art," *J. Modern Power Syst. Clean Energy*, vol. 5, no. 1, pp. 10-19, Jan. 2017.

[17] National Grid and UK Power Networks, "Transmission and distribution interface 2.0 (TDI) - bid document to Ofgem," RIIO Network Innovation Competition (NIC), Tech. Rep. NGET_UKPN_TDI2.0/V.01, 2018. [Online]. Available: https://www.ofgem.gov.uk/ofgem-publications/107804.

[18] SmartNet, "*TSO-DSO coordination for acquiring ancillary services from distribution grids-The SmartNet project final results*", 2019. [Online]. Available: https://www.ofgem.gov.uk/ofgem-publications/107804.

[19] Centrica, "*Cornwall local energy market*", 2018. [Online]. Available: https://www.centrica.com/innovation/cornwall-local-energy-market.

[20] Piclo Flex. [Online]. Available: https://picloflex.com/.

[21] NODES. [Online]. Available: https://nodesmarket.com/.

[22] C. Heinrich, C. Ziras, A. L. Syrri, and H. W. Bindner, "EcoGrid 2.0: A large-scale field trial of a local flexibility market," *Appl. Energy*, vol. 261, pp. 114399, 2020.

[23] European Union. (2021, April 1). *EMPOWER*. [Online]. Available: https://cordis.europa.eu/project/id/646476.

[24] E. Mengelkamp, J. Gärttner, K. Rock, S. Kessler, L. Orsini, C. Weinhardt, "Designing microgrid energy markets: A case study: The Brooklyn microgrid", *Appl. Energy*, vol. 210, pp. 870-880, 2018.

[25] P2P-SmarTest. [Online]. Available: https://www.p2psmartest-h2020.eu/deliverables.

[26] I. J. Perez-Arriaga, J. D. Jenkins, and C. Batlle, "A regulatory framework for an evolving electricity sector: Highlights of the MIT utility of the future study," *Econ. Energy Environ. Policy*, vol. 6, no. 1, 2017, pp. 71-92.

[27] F. Li and R. Bo, "DCOPF-based LMP simulation: algorithm, comparison with ACOPF, and sensitivity." *IEEE Trans. Power Syst.*, vol. 22, no. 4, pp. 1475-1485, Oct. 2007.

[28] P. M. Sotkiewicz and J. M. Vignolo, "Nodal pricing for distribution networks: Efficient pricing for efficiency enhancing DG," *IEEE Trans. Power Syst.*, vol. 21, no. 2, pp. 1013-1014, May 2006.

[29] G. T. Heydt, B. H. Chowdhury *et al.*, "Pricing and control in the next power distribution system," *IEEE Trans. Smart Grid*, vol. 3, no. 2, pp. 907-914, Jun. 2012.

[30] H. Yuan, F. Li *et al.*, "Novel linearized power flow and linearized OPF models for active distribution networks with application in distribution LMP," *IEEE Trans. Smart Grid*, vol. 9, no. 1, pp. 438-448, Jan. 2018.

[31] Y. K. Renani, M. Ehsan, and M. Shahidehpour, "Optimal transactive market operations with distribution system operators," *IEEE Trans. Smart Grid*, vol. 9, no. 6, pp. 6692-6701, Nov. 2018.

[32] W. Wei, *et al.*, "Estimating DLMP confidence intervals in distribution networks with AC power flow model and uncertain renewable generation," *IET Gener. Transm. Distrib.*, vol. 14, no. 8, pp. 1467-1475, Mar. 2020.

[33] I. Alsaleh and L. Fan, "Distribution locational marginal pricing (DLMP) for multiphase systems," *2018 North American Power Symposium (NAPS)*, IEEE, 2018, Fargo, ND, USA, pp. 1-6.

[34] J. Wei *et al.*, "DLMP using three-phase current injection OPF with renewables and demand response," *IET Renew. Power Gener.*, vol. 13, no.7, pp. 1160-1167, May 2019.

[35] B. Wang, N. Tang, R. Bo, and F. Li, "Three-phase DLMP model based on linearized power flow for distribution with application to DER benefit studies," *Int. J. Electr. Power Energy Syst.*, vol. 130, pp. 106884, 2021.

[36] L. Edmonds *et al.*, "Three-phase distribution locational marginal pricing to manage unbalanced variable renewable energy," *IEEE Power Energy Soc. Gen. Meeting*, IEEE, 2020, pp. 1-5.

[37] A. Papavasiliou, "Analysis of distribution locational marginal prices," *IEEE Trans. Smart Grid*, vol. 9, no. 5, pp. 4872-4882, Sep. 2017.

[38] Y. Wu, M. Barati, and G. J. Lim, "A pool strategy of microgrid in power distribution electricity market," *IEEE Trans. Power Syst.*, vol. 35, no. 1, pp. 3-12, Jan. 2019.

[39] J. C. D. Prado and W. Qiao, "A stochastic bilevel model for an electricity retailer in a liberalized distributed renewable energy market," *IEEE Trans. Sustain. Energy*, vol. 11, no. 4, pp. 2803-2812, Oct. 2020.

[40] Federal Energy Regulatory Commission (FERC), Order No. 2222, "A new day for distributed energy resources," Sept. 17, 2020. [Online]. Available: https://www.ferc.gov/media/ferc-order-no-2222-fact-sheet.

[41] "Battery storage in the United States: an update on market trends," U.S. Energy Information Administration, Washington, DC, USA, July 2020.

[42] S. J. Kazempour, A. J. Conejo, and C. Ruiz, "Strategic bidding for a large consumer," *IEEE Trans. on Power Syst.*, vol. 30, no. 2, pp. 848-856, July 2014.

[43] D. S. Kirschen and G. Strbac, "Introduction," in *Fundamentals of Power System Economics*, 2nd ed., Jersey City, John Wiley & Sons, 2018.

[44] X. Kou, F. Li *et al.*, "A scalable and distributed algorithm for managing residential demand response programs using alternating direction method of multipliers (ADMM)," *IEEE Trans. Smart Grid*, vol. 11, no. 6, pp. 4871- 4882, June 2020.

[45] P. Cramton, "Electricity market design," *Oxford Review of Economic Policy*, vol.33, no.4, pp. 589-612, 2017.

[46] R. Mieth and Y. Dvorkin, "Distribution electricity pricing under uncertainty," *IEEE Trans. Power Syst*, vol. 35, no. 3, pp. 2325-2338, May 2020.

[47] L. Wang, Z. Zhu, C. Jiang, and Z. Li, "Bi-level robust optimization for distribution system with multiple microgrids considering uncertainty distribution locational marginal price," *IEEE Trans. Smart Grid*, vol. 12, no. 2, pp. 1104-1117, Mar. 2021.

[48] J. Kim and Y. Dvorkin, "A P2P-dominant distribution system architecture," *IEEE Trans. Power Syst.*, vol. 35, no. 4, pp. 2716-2725, July 2020.

[49] T. Sousa, T. Soares, P. Pinson, F. Moret, T. Baroche, and E. Sorin, "Peer-to-peer and community-based markets: A comprehensive review." *Renew. Sustain. Energy Rev*, vol. 104, pp. 367-378, 2019.

[50] W. Tushar, T. K. Saha, C. Yuen, D. Smith, and H. V. Poor, "Peer-to-peer trading in electricity networks: An overview," *IEEE Trans. Smart Grid*, vol. 11, no. 4, pp. 3185-3200, July 2020.

[51] W. Tushar, C. Yuen, T. K. Saha, T. Morstyn, A. C. Chapman, M. J. E. Alam, S. Hanif, and H. V. Poor, "Peer-to-peer energy systems for connected communities: A review of recent advances and emerging challenges," *Appl. Energy*, vol. 282, pp. 116131, 2021.

[52] Y. Liu, L. Wu, and J. Li, "A two-stage market-oriented peer-to-peer energy trading model for distribution systems with participation of utility," *CSEE J. Power Energy Syst.*, vol. 7, no. 5, pp. 893-902, Sept. 2021.

[53] K. Zhang, S. Troitzsch, S. Hanif, and T. Hamacher, "Coordinated market design for peer-to-peer energy trade and ancillary services in distribution grids," *IEEE Trans. Smart Grid*, vol. 11, no. 4, pp. 2929–2941, July 2020.

[54] H. Haggi and W. Sun, "Multi-round double auction-enabled peer-to-peer energy exchange in active distribution networks," *IEEE Trans. Smart Grid*, vol. 12, no. 5, pp. 4403–4414, Sept. 2021.

[55] E. Münsing, J. Mather, and S. Moura, "Blockchains for decentralized optimization of energy resources in microgrid networks," *Proc. IEEE Conf. Control Technol. Appl.*, Kohala Coast, Hawaii, USA, 2017, pp. 2164–2171.

[56] M. Shahidehpour, M. Yan, P. Shikhar, S. Bahramirad, and A. Paaso "Blockchain for Peer-to-Peer transactive energy trading in networked microgrids: Providing an effective and decentralized strategy." *IEEE Electrif. Mag.*, vol. 8, no.4, pp. 80-90, Dec. 2020.

[57] F. Xin *et al.*, "Introducing uncertainty components in locational marginal prices for pricing wind power and load uncertainties," *IEEE Trans. Power Syst.*, vol. 34, no. 3, pp. 2013-2024, Jan. 2019.

[58] P. Andrianesis and M Caramanis, "Distribution network marginal costs: enhanced AC OPF including transformer degradation," *IEEE Trans. Smart Grid*, vol. 11, no. 5, pp. 3910-3920, Sept. 2020.

[59] T. Yang, Y. Guo, L. Deng, H. Sun, and W. Wu, "A linear branch flow model for radial distribution networks and its application to reactive power optimization and network reconfiguration," *IEEE Trans. Smart Grid*, vol. 12, no. 3, pp. 2027-2036, May 2021.

[60] M. E. Baran and F. F. Wu, "Network reconfiguration in distribution systems for loss reduction and load balancing," *IEEE Trans. Power Del.*, vol. 4, no. 2, pp. 1401-1407, Apr. 1989.

[61] C. Zhang, Y. Xu, Z. Dong, and J. Ravishankar, "Three-stage robust inverter-based voltage/var control for distribution networks with high-level PV," *IEEE Trans. Smart Grid*, vol. 10, no. 1, pp. 782-793, Jan. 2019.

[62] L. Gan and S. H. Low, "Convex relaxations and linear approximation for optimal power flow in multiphase radial networks," *2014 Power Systems Computation Conference*, Wroclaw, Poland, 2014, pp. 1-9.

[63] Y. Du, F. Li, *et al.*, "A cooperative game approach for coordinating multi-microgrid operation within distribution systems," *Appl. Energy*, vol. 222, pp. 383-395, 2018.

[64] B. S. K. Patnam and N. M. Pindoriya, "DLMP calculation and congestion minimization with EV aggregator loading in a distribution network using bilevel program," *IEEE Systems Journal*, vol. 15, no. 2, pp. 1835-1846, June 2021.

[65] J. Zhang, M. Cui, B. Li, H. Fang, and Y. He, "Fast solving method based on linearized equations of branch power flow for coordinated charging of EVs (EVCC)," *IEEE Trans. Veh. Technol.*, vol. 68, no. 5, pp. 4404–4418, May 2019.

[66] B. Park, Y. Chen, F. Li *et al.*, "Optimal demand response incorporating distribution LMP (DLMP) with PV generation uncertainty," *IEEE Trans. Power Syst.*, vol. 37, no. 2, pp. 982-995, Mar. 2022.

[67] Y. Liu, N. Zhang, Y. Wang, J. Yang, and C. Kang, "Data-driven power flow linearization: A regression approach," *IEEE Trans. Smart Grid*, vol. 10, no. 3, pp. 2569-2580, May 2019.

[68] Y. Liu, Y. Wang, N. Zhang, D. Lu, and C. Kang, "A data-driven approach to linearize power flow equations considering measurement noise," *IEEE Trans. Smart Grid*, vol. 11, no. 3, pp. 2576-2587, May 2020.

[69] B. Chen, C. Chen, J. Wang, and K. L. Butler-Purry, "Sequential service restoration for unbalanced distribution systems and microgrids," *IEEE Trans. Power Syst.*, vol. 33, no. 2, pp. 1507-1520, Mar. 2018.

[70] Y. Wang, N. Zhang, H. Li, J. Yang, and C. Kang, "Linear three-phase power flow for unbalanced active distribution networks with PV nodes," *CSEE J. Power Energy Syst.*, vol. 3, no. 3, pp. 321-324, Sep. 2017.

[71] X. Bai, H. Wei, K. Fujisawa, and Y. Wang, "Semidefinite programming for optimal power flow problems," *Int. J. Elect. Power Energy Syst.*, vol. 30, pp. 383-392, 2008.

[72] J. Lavaei and S. H. Low, "Zero duality gap in optimal power flow problem," *IEEE Trans. Power Syst.*, vol. 27, no. 1, pp. 92-107, Feb. 2012.

[73] C. Zhao, E. Dall'Anese, and S. H. Low, "Convex relaxation of OPF in multiphase radial networks with delta connection," *Proc. of Bulk Power Systems Dynamics and Control Symposium*, Charleston, SC, USA, 2017, pp. 0885-8950.

[74] Y. Liu, J. Li, and L. Wu, "Distribution system restructuring: Distribution LMP via unbalanced ACOPF," *IEEE Trans. Smart Grid*, vol. 9, no. 5, pp. 4038-4048, Sept. 2018.

[75] R. A. Jabr, "Radial distribution load flow using conic programming," *IEEE Trans. Power Syst.*, vol. 21, no. 3, pp. 1458-1459, Aug. 2006.

[76] M. Farivar and S. H. Low, "Branch flow model: Relaxations and convexification–Part I," *IEEE Trans. Power Syst.*, vol. 28, no. 3, pp. 2554–2572, Aug. 2013.

[77] W. Wei, J. Wang, N. Li, and S. Mei, "Optimal power flow of radial networks and its variations: A sequential convex optimization approach," *IEEE Trans. Smart Grid*, vol. 8, no. 6, pp. 2974-2987, Nov. 2017.

[78] S. H. Low, "Convex relaxation of optimal power flow–Part I: Formulations and equivalence," *IEEE Trans. Control Netw. Syst.*, vol. 1, no. 1, pp. 15–27, Mar. 2014.

[79] S. H. Low, "Convex relaxation of optimal power flow–Part II: Exactness," *IEEE Trans. Control Netw. Syst.*, vol. 1, no. 2, pp. 177–189, June 2014.

[80] S. Nematshahi and H. R. Mashhadi, "Distribution network reconfiguration with the application of DLMP using genetic algorithm," *Proc. IEEE Electr. Power Energy Conf. (EPEC)*, Saskatoon, Saskatchewan, Canada, 2017, pp. 1-5.

[81] R. Li, Q. Wu, and S. S. Oren, "Distribution locational marginal pricing for optimal electric vehicle charging management," *IEEE Trans. Power Syst.*, vol. 29, no. 1, pp. 203-211, Jan. 2014.

[82] S. Huang, Q. Wu, S. S. Oren, R. Li, and Z. Liu, "Distribution locational marginal pricing through quadratic programming for congestion management in distribution networks," *IEEE Trans. Power Syst.*, vol. 30, no. 4, pp. 2170-2178, Jul. 2015.

[83] W. Liu, *et al.*, "Day-ahead congestion management in distribution systems through household demand response and distribution congestion prices," *IEEE Trans. Smart Grid*, vol. 5, no. 6, pp. 2739–2747, Jul. 2014.

[84] X. Wang, F. Li, et al., "Tri-level scheduling model considering residential demand flexibility of aggregated HVACs and EVs under distribution LMP (DLMP)," *IEEE Trans. Smart Grid*, vol. 12, no. 5, pp. 3990-4002, Sept. 2021.

[85] M. Caramanis, E. Ntakou, W. Hogan, A. Chakrabortty, and J. Schoene, "Co-optimization of power and reserves in dynamic T&D power markets with nondispatchable renewable generation and distributed energy resources," *Proc. IEEE*, vol. 104, no. 4, pp. 807–836, May 2016.

[86] Z. Yuan, M. R. Hesamzadeh, and D. R. Biggar, "Distribution locational marginal pricing by convexified ACOPF and hierarchical dispatch," *IEEE Trans. Smart Grid*., vol. 9, no. 4, pp. 3133-3142, July 2018.

[87] P. Jacquot, "DLMP-based coordination procedure for decentralized demand response under distribution network constraints," *arXiv preprint arXiv:2004.06004*, 2020.

[88] Z. Yang, H. Zhong, A. Bose, T. Zheng, Q. Xia, and C. Kang, "A linearized OPF model with reactive power and voltage magnitude: A pathway to improve the MW-only DC OPF," *IEEE Trans. Power Syst.*, vol. 33, no. 2, pp. 1734-1745, Mar. 2018.

[89] H. Ye, Y. Ge, M. Shahidehpour, and Z. Li, "Uncertainty marginal price, transmission reserve, and day-ahead market clearing with robust unit commitment," *IEEE Trans. Power Syst.*, vol. 32, no. 3, pp. 1782-1795, May 2017.

[90] L. Exizidis, J. Kazempour, A. Papakonstantinou, P. Pinson, Z. De Grève, and F. Vallée, "Incentive-compatibility in a two-stage stochastic electricity market with high wind power penetration," *IEEE Trans. Power Syst.*, vol. 34, no. 4, pp. 2846-2858, July 2019.

[91] H. Zhong, L. Xie, and Q. Xia, "Coupon incentive-based demand response: Theory and case study," *IEEE Trans. Power Syst.*, vol. 28, no. 2, pp. 1266-1276, May 2013.

[92] A. D. Giorgio and L. Pimpinella, "An event driven smart home controller enabling consumer economic saving and automated demand side management," *Appl. Energy*, vol. 96, pp. 92-103, 2012.

[93] K. Herter and S. Wayland, "Residential response to critical-peak pricing of electricity: California evidence," *Energy*, vol. 35, pp. 1561-1567, 2010.

[94] J. Vuelvas and F. Ruiz, "Rational consumer decisions in a peak time rebate program," *Electric Power Systems Research*, vol. 143, pp. 533-543, 2017.

[95] S. Bhattacharya, K. Kar and J. H. Chow, "Optimal precooling of thermostatic loads under time-varying electricity prices," *Proc. IEEE American Control Conference (ACC)*, Seattle, WA, USA, 2017, pp. 1407-1412.

[96] Y. J. Kim, "Optimal price based demand response of HVAC systems in multizone office buildings considering thermal preferences of individual occupants building," *IEEE Trans. Ind. Informat.*, vol. 14, no. 11, pp. 5060-5073, Nov. 2018.

[97] W. Zhong, K. Xie, Y. Liu, S. Xie, and L. Xie, "Nash mechanisms for market design based on distribution locational marginal prices," *IEEE Trans. Power Syst.*, early access, 2022.

[98] W. Tushar, T. K. Saha, C. Yuen, P. Liddell, R. Bean, and H. V. Poor, "Peer-to-peer energy trading with sustainable user participation: A game theoretic approach," *IEEE Access*, vol. 6, pp. 62932-62943, Oct. 2018.

[99] W. Tushar, T. K. Saha, C. Yuen, T. Morstyn, and M. D. McCulloch, "A motivational game-theoretic approach for peer-to-peer energy trading in the smart grid," *Appl. Energy*, vol. 243, pp. 10-20, June 2019.

[100] C. Long, C. Z. J. Wu, L. Thomas, M. Cheng, and N. Jenkins, "Peer-to-

peer energy trading in a community microgrid," in Proc. *IEEE Power Energy Soc. Gen. Meeting,* Chicago, IL, USA, Jul. 2017, pp. 2-6.

[101] Y. Zhou, J. Wu, and C. Long, "Evaluation of peer-to-peer energy sharing mechanisms based on a multiagent simulation framework," *Appl. Energy,* vol. 222, pp. 993-1022, July 2018.

[102] J. Li, Y. Ye, D. Papadaskalopoulos, and G. Strbac, "Computationally efficient pricing and benefit distribution mechanisms for incentivizing stable peer-to-peer energy trading," *IEEE Internet Things J.,* vol. 8, no. 2, pp. 734-749, Jan. 2021.

[103] H. Wang and J. Huang, "Incentivizing energy trading for interconnected microgrids," *IEEE Trans. Smart Grid,* vol. 9, no. 4, pp. 2647-2657, Jun. 2018.

[104] W. Zhong, S. Xie, K. Xie, Q. Yang, and L. Xie, "Cooperative P2P energy trading in active distribution networks: An MILP-based Nash bargaining solution," *IEEE Trans. Smart Grid,* vol. 12, no. 2, pp. 1264-1276, Mar. 2021.

[105] Y. Wang, S. Wang, and L. Wu, "Distributed optimization approaches for emerging power systems operation: A review," *Electric Power Systems Research,* vol. 144, pp. 127-135, 2017.

[106] S. Hanif, K. Zhang, C. M. Hackl, M. Barati, H. B. Gooi, and T. Hamacher, "Decomposition and equilibrium achieving distribution locational marginal prices using trust-region method," *IEEE Trans. Smart Grid.,* vol. 10, no. 3, pp. 3269-3281, May 2019.

[107] S. Suganya, S. C. Raja, and P. Venkatesh, "Simultaneous coordination of distinct plug-in hybrid electric vehicle charging stations: A modified particle swarm optimization approach," *Energy,* vol. 138, pp. 92-102, 2017.

[108] D. K. Molzahn *et al.*, "A survey of distributed optimization and control algorithms for electric power systems," *IEEE Trans. Smart Grid,* vol. 8, no. 6, pp. 2941-2962, Nov. 2017.

[109] S. Boyd, N. Parikh, E. Chu, B. Peleato, and J. Eckstein, "Distributed optimization and statistical learning via the alternating direction method of multipliers," *Found. Trends Mach. Learn.,* vol. 3, no. 1, pp. 1-122, 2011.

[110] X. Kou, F. Li *et al.*, "A comprehensive scheduling framework using SP-ADMM for residential demand response with weather and consumer uncertainties," *IEEE Trans. Power Syst.,* vol. 36, no. 4, pp. 3004-3016, July 2021.

[111] Y. Wang, L. Wu, and S. Wang, "A fully-decentralized consensus-based ADMM approach for DC-OPF with demand response," *IEEE Trans. Smart Grid,* vol. 8, no. 6, pp. 2637-2647, Mar. 2016

[112] Z. Tan, P. Yang, and A. Nehorai, "An optimal and distributed demand response strategy with electric vehicles in the smart grid," *IEEE Trans. Smart Grid,* vol. 5, no. 2, pp. 861-869, Mar. 2014.

[113] D. H. Nguyen, S. Azuma, and T. Sugie, "Novel control approaches for demand response with real-time pricing using parallel and distributed consensus-based ADMM," *IEEE Trans. Ind. Electron.,* vol. 66, no. 10, pp. 7935-7945, Dec. 2018.

[114] F. Shen, Q. Wu, X. Jin, B. Zhou, C. Li, and Y. Xu, "ADMM-based market clearing and optimal flexibility bidding of distribution-level flexibility market for day-ahead congestion management of distribution networks," *Int. J. Electr. Power Energy Syst.,* vol.123, pp.1-14, 2020.

[115] Y. He, Q. Chen, J. Yang, Y. Cai, and X. Wang, "A multi-block ADMM based approach for distribution market clearing with distribution locational marginal price," *Int. J. Electr. Power Energy Syst.,* vol. 128, pp. 106635, 2021.

[116] K. Zhang, S. Hanif, C. M. Hackl, and T. Hamacher, "A framework for multi-regional real-time pricing in distribution grids," *IEEE Trans. Smart Grid,* vol. 10, no. 6, pp. 6826-6838, Nov. 2019.

[117] H. M. Kim *et al.*, "Target cascading in optimal system design," *J. Mech. Des.* vol. 125, no. 3, pp. 474-480, Sept. 2003.

[118] A. Kargarian *et al*, "Toward distributed/decentralized DC optimal power flow implementation in future electric power systems," *IEEE Trans. on Smart Grid,* vol. 9, no. 4, pp.2574-2594. Oct. 2016.

[119] M. Xie *et al.*, "Autonomous optimized economic dispatch of active distribution system with multi-microgrids," *Energy,* vol. 153, pp. 479-489, 2018.

[120] P. Du *et al.*, "A bi-level linearized dispatching model of active distribution network with multi-stakeholder participation based on analytical target cascading," *IEEE Access,* vol. 7, pp. 154844-154858, Oct. 2019.

[121] A. Mohammadi and A. Kargarian, "Accelerated and robust analytical target cascading for distributed optimal power flow," *IEEE Trans. Ind. Informat.,* vol. 16, no. 12, pp. 7521-7531, Feb. 2020.

[122] J. Zhang, L. Che, X. Wan, and M. Shahidehpour, "Distributed hierarchical coordination of networked charging stations based on peer-to-peer trading and EV charging flexibility quantification," *IEEE Trans. Power Syst.,* early access, 2021.

[123] M. Kraning, E. Chu, J. Lavaei, and S. Boyd, "Dynamic network energy management via proximal message passing," *Found. Trends Optim.,* vol. 1, no. 2, pp. 70-122, 2013.

[124] E. Ntakou and M. Caramanis, "Distribution network electricity market clearing: Parallelized PMP algorithms with minimal coordination," *Proc. IEEE 53rd Annu. Conf. Decis. Control (CDC),* Los Angeles, CA, USA, 2014, pp. 1687-1694.

[125] E. Ntakou and M. Caramanis, "Enhanced convergence rate of inequality constraint shadow prices in PMP algorithm cleared distribution power markets," *Proc. Amer. Control Conf.,* Boston, MA, USA, 2016, pp.1433-1439.

[126] S. Chakrabarti and R. Baldick, "Look-ahead SCOPF (LASCOPF) for tracking demand variation via auxiliary proximal message passing (APMP) algorithm," *Int. J. Electr. Power Energy Syst.,* vol. 116, pp. 105533, 2020.

[127] G. Cohen, "Auxiliary problem principle and decomposition of optimization problems," *J. Optim. Theory Appl.,* vol. 32, no. 3, pp. 277–305, Nov. 1980.

[128] A. Ahmadi-Khatir, A. J. Conejo, and R. Cherkaoui, "Multi-area unit scheduling and reserve allocation under wind power uncertainty," *IEEE Trans. Power Syst.,* vol. 29, no. 4, pp. 1701–1710, Jul. 2014.

[129] M. Alkhraijah, M. Alowaifeer, S. Grijalva, and D. K. Molzahn, "Distributed multi-period DCOPF via an auxiliary principle problem algorithm," *IEEE Tex. Power Energy Conf.*, College Station, TX, USA, 2021, pp. 1-6.

[130] A. R. Di Fazio, C. Risi, M. Russo, and M. De Santis, "Distributed voltage optimization based on the auxiliary problem principle in active distribution systems," *55th Int. Univ. Power Eng. Conf.*, Torino, Italy, 2020, pp. 1-6.

[131] C. Eid, P. Codani, Y. Perez, J. Reneses, and R. Hakvoort, "Managing electric flexibility from distributed energy resources: A review of incentives for market design," *Renew. Sustain. Energy Rev.,* vol. 64, pp. 237-247, 2016.

[132] T. Zhao and Z. Ding, "Cooperative optimal control of battery energy storage system under wind uncertainties in a microgrid," *IEEE Trans. Power Syst.,* vol. 33, no. 2, pp. 2292-2300, Mar. 2018.

[133] P. Siano and D. Sarno, "Assessing the benefits of residential demand response in a real time distribution energy market," *Appl. Energy,* vol. 161, pp. 533-551, 2016.

[134] H. Mohsenian-Rad, "Optimal demand bidding for time-shiftable loads," *IEEE Trans. Power Syst.,* vol. 30, no. 2, pp. 939-951, Mar. 2015.

[135] S. Rahman, A. Haque, and Z. Jing, "Modeling and performance evaluation of grid-interactive efficient buildings (GEB) in a microgrid environment," *IEEE Open Access Journal of Power and Energy,* vol. 8, pp. 423-432, 2021.

[136] M. Song, W. Sun, Y. Wang, M. Shahidehpour, Z. Li, and C. Gao, "Hierarchical scheduling of aggregated TCL flexibility for transactive energy in power systems," *IEEE Trans. Smart Grid,* vol. 11, no. 3, pp. 2452-2463, May 2020.

[137] Y. Du, F. Li, K. Kurte, J. Munk, and H. Zandi, "Demonstration of intelligent HVAC load management with deep reinforcement learning," *IEEE Power Energy Mag.,* vol. 20, no. 3, pp. 42-53, May-June 2022.

[138] J. Wei, Y. Zhang, J. Wang, and L. Wu, "Distribution LMP-based demand management in industrial park via a bi-level programming approach," *IEEE Trans. Sustain. Energy,* vol. 12, no. 3, pp. 1695-1706, July 2021.

[139] Y. Du and F. Li, "A hierarchical real-time balancing market considering multi-microgrids with distributed sustainable resources," *IEEE Trans. Sustain. Energy,* vol. 11, no.1, pp. 72-83. Jan. 2018.

[140] Z. Wang *et al.*, "Robust optimization based optimal DG placement in microgrids," *IEEE Trans. Smart Grid,* vol. 5, no. 5, pp. 2173-2182, Sept. 2014.

[141] Y. Tang and S. H. Low, "Optimal placement of energy storage in distribution networks," *IEEE Trans. Smart Grid,* vol. 8, no. 6, pp. 3094-3103, Nov. 2017.

[142] L. Bai, T. Jiang, F. Li, *et al*, "Distributed energy storage planning in soft open point based active distribution networks incorporating network reconfiguration and DG reactive power capability," *Appl. Energy,* vol. 210, pp. 1082-1091, 2018.

[143] M. R. Jannesar, A. Sedighi, M. Savaghebi, and J. M. Guerrero, "Optimal placement, sizing, and daily charge/discharge of battery energy storage in low voltage distribution network with high photovoltaic penetration", *Appl. Energy,* vol. 226, pp. 957-966, 2018.

[144] X. Su, Z. Zhang, Y. Liu, Y. Fu, F. Shahnia, C. Zhang, and Z. Dong, "Sequential and comprehensive BESSs placement in unbalanced active distribution networks considering the impacts of BESS dual attributes on sensitivity," *IEEE Trans. Power Syst.,* vol. 36, no. 4, pp. 3453-3464, July 2021.

[145] M. H. S. Boloukat and A. A. Foroud, "Multiperiod planning of distribution networks under competitive electricity market with penetration of several microgrids, part I: Modeling and solution methodology," *IEEE Trans. Ind. Informat.*, vol. 14, no. 11, pp. 4884-4894, Nov. 2018.

[146] R. Atia and N. Yamada, "Distributed renewable generation and storage systems sizing in deregulated energy markets," *2015 International Conference on Renewable Energy Research and Applications (ICRERA)*, Palermo, Italy, Nov. 2015, pp. 258-262.

[147] E. Samani and F. Aminifar, "Tri-level robust investment planning of DERs in distribution networks with AC constraints," *IEEE Trans. Power Syst.*, vol. 34, no. 5, pp. 3749-3757, Sept. 2019.

[148] P. Lamaina, D. Sarno, P. Siano, A. Zakariazadeh, and R. Romano, "A model for wind turbines placement within a distribution network acquisition market," *IEEE Trans. Ind. Informat.*, vol. 11, no. 1, pp. 210-219, Feb. 2015.

[149] X. Xiao, F. Wang, M. Shahidehpour, Z. Li, and M. Yan, "Coordination of distribution network reinforcement and DER planning in competitive market," *IEEE Trans. Smart Grid*, vol. 12, no. 3, pp. 2261-2271, May 2021.

[150] X. Wang, F. Li, Q. Zhang, Q. Shi, and J. Wang, "Profit-oriented BESS siting and sizing in deregulated distribution systems," *IEEE Trans. Smart Grid*, early access, 2022.

[151] E. A. Farsani, H. A. Abyaneh, M. Abedi, and S. H. Hosseinian, "A novel policy for LMP calculation in distribution networks based on loss and emission reduction allocation using nucleolus theory," *IEEE Trans. Smart Grid*, vol. 31, no. 1, pp. 143-152, Jan. 2016.

[152] S. Hanif, T. Massier, H. B. Gooi, T. Hamacher, and T. Reindl, "Cost Optimal Integration of Flexible Buildings in Congested Distribution Grids," *IEEE Trans. Power Syst.*, vol. 32, no. 3, pp. 2254-2266, May 2017.

[153] J. Hu, J. Wu, X. Ai, and N. Liu, "Coordinated energy management of prosumers in a distribution system considering network congestion," *IEEE Trans. Smart Grid*, vol. 12, no. 1, pp. 468-478, Jan. 2021.

[154] CAISO. (2021, Sep. 8). *Flex alert extended through tomorrow due to widespread heat in west*. [Online]. Available: https://www.flexalert.org/news.

[155] H. Chen, L. Fu, L. Bai, *et al*, "Distribution market-clearing and pricing considering coordination of DSOs and ISO: An EPEC approach", *IEEE Trans. Smart Grid*, vol. 12, no. 4, pp. 3150-3162, July 2021.

[156] R. Kravitz, "Coordination of transmission and distribution operations in a high distributed energy resource electric grid," California Energy Commission, Sacramento, CA, USA, June 2017.

[157] Y. Shen, Y. Li, Q. Zhang, F. Li, and Z. Wang, "Consumer psychology based optimal portfolio design for demand response aggregators," *J. Mod. Power Syst. Clean Energy*, vol. 9, no. 2, pp. 431-439, Mar. 2021.

[158] S. Hanif, P. Creutzburg, H. B. Gooi, and T. Hamacher, "Pricing mechanism for flexible loads using distribution grid hedging rights," *IEEE Trans. Power Syst.*, vol. 34, no. 5, pp. 4048-4059, Sept. 2019.

[159] A. K. Zarabie, S. Das, and M. N. Faqiry, "Fairness-regularized DLMP-based bilevel transactive energy mechanism in distribution systems," *IEEE Trans. Smart Grid*, vol. 10, no. 6, pp. 6029-6040, Jan. 2019.

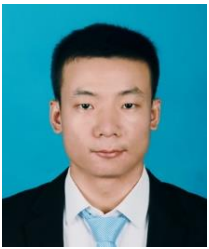

**Xiaofei Wang** (S'19) received the B.S.E.E. degree from North China Electric Power University in 2014, and the M.S.E.E. degree from Wuhan University, China, in 2017. He is currently pursuing the Ph.D. degree in electrical engineering at the University of Tennessee, Knoxville, TN, USA. His research interests include distribution-level market, power system optimization, and demand response.

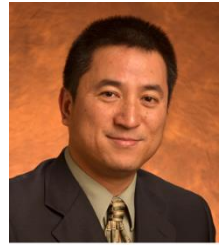

**Fangxing Li** (Fellow, IEEE) is also known as Fran Li. He received the B.S.E.E. and M.S.E.E. degrees from Southeast University, Nanjing, China, in 1994 and 1997, respectively, and the Ph.D. degree from Virginia Tech, Blacksburg, VA, USA, in 2001. Currently, he is the James W. McConnell Professor in electrical engineering and the Campus Director of CURENT at the University of Tennessee, Knoxville, TN, USA. His current research interests include renewable energy integration, demand response, distributed generation and microgrid, energy markets, and power system computing. He was the Past Chair (2020-2021) of IEEE PES Power System Operation, Planning and Economics (PSOPE) Committee. Since 2020, he has been serving as the Editor-In-Chief of *IEEE Open Access Journal of Power and Energy (OAJPE)*.

Dr. Li has received a number of awards and honors including R&D 100 Award in 2020, IEEE PES Technical Committee Prize Paper award in 2019, 3 best paper awards at international journals, and 6 best papers/posters at international conferences.

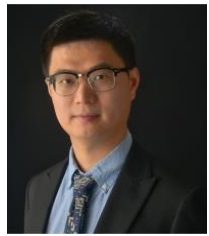

**Linquan Bai** (S'13–M'17–SM'20) received the B.S. and M.S. degrees in electrical engineering from Tianjin University, Tianjin, China, in 2010 and 2013, respectively, and the Ph.D. degree in electrical engineering from the University of Tennessee, Knoxville, TN, USA, in 2017. He is an Assistant Professor with the University of North Carolina at Charlotte, Charlotte, NC, USA. His research interests include grid integration of distributed energy resources, power system modeling and optimization, and electricity markets. He is an Associate Editor of IEEE Transactions on Sustainable Energy, IEEE Power Engineering Letters, and Journal of Modern Power System and Clean Energy.

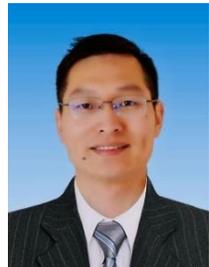

**Xin Fang** (SM'18) received the B.S. degree from Huazhong University of Science and Technology, China, in 2009, the M.S. degree from China Electric Power Research Institute, China, in 2012, and the Ph.D. degree at the University of Tennessee (UT), Knoxville, TN, USA, in 2016. He is currently a senior researcher at National Renewable Energy Laboratory (NREL). Before joining NREL, he was a power system engineer at GE Grid Solutions from 2016 to 2017. Dr. Fang is an associate editor of IEEE Transactions on Sustainable Energy. His research interests include power system planning and optimization, electricity market operation considering renewable energy integration, and cyber-physical transmission and distribution modeling and simulation.